\documentclass[prd,preprint,a4paper,showpacs,nofootinbib,superscriptaddress]{revtex4-1}
\usepackage{bm}
\usepackage{indentfirst}
\usepackage{amsmath}
\usepackage{graphicx}
\usepackage{float}
\usepackage{amssymb}
\usepackage{subfigure}
\usepackage{amssymb}
\usepackage{psfrag}
\usepackage{hyperref}
\usepackage{epstopdf}
\hypersetup{
	colorlinks=true,
	linkcolor=red,
	citecolor=blue,
}

\begin{document}
\title{Frequency response of time-delay interferometry for space-based gravitational wave antenna} 	

\author{Chunyu Zhang}
\email{chunyuzhang@hust.edu.cn}
\affiliation{School of Physics, Huazhong University of Science and Technology,
Wuhan, Hubei 430074, China}

\author{Qing Gao}
\email{gaoqing1024@swu.edu.cn}
\affiliation{School of Physical Science and Technology, Southwest University, Chongqing 400715, China}

\author{Yungui Gong}
\email{Corresponding author. yggong@hust.edu.cn}
\affiliation{School of Physics, Huazhong University of Science and Technology,
Wuhan, Hubei 430074, China}

\author{Dicong Liang}
\email{dcliang@hust.edu.cn}
\affiliation{School of Physics, Huazhong University of Science and Technology,
Wuhan, Hubei 430074, China}

\author{Alan J. Weinstein}
\email{ajw@ligo.caltech.edu}
\affiliation{LIGO Laboratory, California Institute of Technology, Pasadena, California 91125, USA }

\author{Chao Zhang}
\email{chao\_zhang@hust.edu.cn}
\affiliation{School of Physics, Huazhong University of Science and Technology,
Wuhan, Hubei 430074, China}

\begin{abstract}
Space-based gravitational wave detectors cannot keep rigid structures and  precise arm length equality,
so the precise equality of detector arms which is required in a ground-based interferometer to cancel the overwhelming laser noise is impossible.
The time-delay interferometry  method is applied to unequal arm lengths to cancel the laser frequency noise.
We give analytical formulas of the averaged response functions for tensor, vector, breathing and longitudinal polarizations in different TDI combinations,
and obtain their asymptotic behaviors.
At low frequencies, $f\ll f_*$,  the averaged response functions of all TDI combinations increase as $f^2$ for all six polarizations.
The one exception is that the averaged response functions of $\zeta$ for all six polarizations increase as $f^4$ in the equilateral-triangle case.
At high frequencies, $f\gg f_*$, the averaged response functions of all TDI combinations for the tensor and breathing modes fall off as $1/f^2$,
the averaged response functions of all TDI combinations for the vector mode fall off as $\ln(f)/f^2$ ,
and the averaged response functions of all TDI combinations for the longitudinal mode fall as $1/f$.
We also give LISA and TianQin sensitivity curves in different TDI combinations for tensor, vector, breathing and longitudinal polarizations.
\end{abstract}
\maketitle

\section{Introduction}

Gravitational waves (GWs) are disturbances of space-time predicted by Einstein's general relativity. In Einstein's theory
of general relativity, GWs propagate at the speed of
light with two transverse polarization states. In alternative theories of
gravity, GWs may have up to six polarizations and the propagation speed
may differ from the speed of light \cite{Eardley:1974nw,Liang:2017ahj,Hou:2017bqj,Gong:2017bru,Gong:2017kim,Gong:2018cgj,Gong:2018ybk,
Gong:2018vbo,Hou:2018djz,Hou:2018djz}.
The detections of GWs by the Laser Interferometer Gravitational-Wave 
Observatory (LIGO) Scientific Collaboration and the Virgo Collaboration
opened a new window to test general relativity and probe the nature of gravity in the strong field regime \cite{Abbott:2016blz,Abbott:2016nmj,
Abbott:2017vtc,Abbott:2017oio,TheLIGOScientific:2017qsa,
Abbott:2017gyy,LIGOScientific:2018mvr}.
The ground-based detectors, such as Advanced LIGO \cite{Harry:2010zz,TheLIGOScientific:2014jea}, Advanced Virgo \cite{TheVirgo:2014hva} and KAGRA \cite{Somiya:2011np,Aso:2013eba}, operate in the high frequency band (10-$10^4$ Hz). However, there are many important gravitational wave (GW) sources
such as coalescing galactic binaries, coalescing supermassive black hole binaries, and secondary GWs from ultra-slow-roll inflation \cite{Gong:2017qlj,Yi:2017mxs} emitting GWs
with low frequency (mHz-1Hz). The detection of low frequency GWs will help address numerous astrophysical, cosmological, and theoretical problems.
The proposed space-based detectors such as LISA \cite{Danzmann:1997hm,Audley:2017drz},
TianQin \cite{Luo:2015ght}, and TaiJi \cite{Hu:2017mde} probe GWs in the frequency band of millihertz, while DECIGO  \cite{Kawamura:2011zz} operates in the frequency band of 0.1 to 10 Hz.

For ground-based interferometric GW detectors, we do not need to worry about the frequency dependence of the antenna response because the wavelength of in-band GWs is larger than
the arm length of the detector. For space-based interferometric GW detectors, the distance between spacecraft (SC) is comparable or even larger than the wavelength of in-band GWs
and it is impossible to maintain the precise equality of the arm lengths. 
Since in the interferometric measurements,
laser frequency noise dominates the expected GW signals by several orders of magnitude,
it experiences different time delays in the arms and hence will not cancel out when the beams are recombined.
Time-delay interferometry (TDI) proposed in \cite{Tinto:1999yr,Armstrong_1999} is a technique applied to unequal arm lengths that significantly reduces this laser frequency noise.
By synthesizing virtual equal arm interferometric measurements with TDI, the laser frequency noise was brought below the GW signals \cite{Estabrook:2000ef}.
Furthermore, a rigorous and systematic procedure based on algebraic geometrical methods and commutative algebra was proposed to cancel the laser frequency noise \cite{Dhurandhar:2002zcl}.
In the first generation of TDI, a static array is assumed. The second generation of TDI applies to a rotating and flexing configuration with arm lengths varying linearly in time.
Recently, the effects of the flexing-filtering coupling on the TDI residual laser noise for LISA was considered \cite{Bayle:2018hnm}.
In \cite{Tinto:2018kij}, the authors discussed the clock-noise calibration of the phase fluctuations of the onboard ultrastable oscillators for the second generation formulation of TDI.
For more discussion on TDI algorithm and its application to LISA,
please see \cite{Tinto:2001ii,Tinto:2001ui,Hogan:2001jn,Tinto:2002de,Prince:2002hp,Shaddock:2003dj,
Tinto:2003vj,Tinto:2003uk,Cornish:2003tz,Nayak:2003na,Armstrong:2003ut,Tinto:2004nz,
Vallisneri:2004bn,RajeshNayak:2004jzp,Tinto:2004wu,Nayak:2005un,Romano:2006rj,Tinto:2014lxa,Wang:2017aqq}
and references therein.

In this paper, we discuss the frequency response of TDI combinations for space-based GW interferometers by averaging the transfer functions over source directions and polarizations. For equal arm space-based interferometric detectors without
optical cavities, an analytical formula for the averaged response function
of the tensor mode was derived in \cite{Larson:1999we} and the generalization
to other polarizations was obtained in \cite{Liang:2019pry}.
The averaged response function for all six possible polarization was
also discussed in \cite{Blaut:2012zz} by numerical simulation.
For LISA, the averaged response functions of TDI combinations
for all six possible polarization
were obtained with numerical simulation in \cite{Tinto:2010hz}.
An analytical formula for the averaged response function
of the tensor mode for unequal arm space-based GW detectors was derived in \cite{Larson:2002xr}. The purpose of this paper is to
extend the discussions on the averaged response functions in \cite{Larson:1999we,Larson:2002xr,Liang:2019pry}
to different TDI combinations for all polarizations so that analytical expressions are derived to understand the asymptotic behaviors of the averaged response functions.

The paper is organized as follows. In Sec. \ref{sec2} we discuss the antenna response functions of TDI combinations.
The analytical formulas for averaged response functions of different TDI combinations
are derived and the asymptotic behaviors of the averaged response function
are analyzed in Sec. \ref{sec3}. The discussion on the averaged response functions
for equal arm space-based interferometric GW detectors to massive gravitons
and the analytical formulas for different TDI combinations are presented in Appendixes \ref{appa} and \ref{appb}. In Sec. \ref{sec4},
we give the sensitivity curves for LISA and TianQin.
The paper is concluded in Sec. \ref{sec5}.

\section{Antenna response functions}
\label{sec2}
\subsection{Polarization tensors}

For a GW propagating along the direction $\hat{\Omega}(\theta,\phi)$, we introduce two perpendicular unit vectors $\hat{p}$ and $\hat{q}$ to from the orthonormal coordinate system such that $\hat{\Omega}=\hat{p}\times\hat{q}$. To account for the rotational degree of freedom around $\hat{\Omega}$, we introduce the polarization angle $\psi$ to form two new orthonormal vectors,
\begin{equation}
	\hat{r}=\cos\psi \hat{p}+\sin\psi\hat{q},\qquad\hat{s}=-\sin\psi\hat{p}+\cos\psi\hat{q}.
\end{equation}
With the orthonormal basis, the six polarization tensors are defined as
\begin{equation}
\begin{split}
	e^+_{ij}=\hat{r}_i\hat{r}_j-\hat{s}_i\hat{s}_j, \qquad & e^\times_{ij}=\hat{r}_i\hat{s}_j+\hat{s}_i\hat{r}_j, \\
	e^x_{ij}=\hat{r}_i\hat{\Omega}_j+\hat{\Omega}_i\hat{r}_j, \qquad & e^y_{ij}=\hat{s}_i\hat{\Omega}_j+\hat{\Omega}_i\hat{s}_j, \\
	e^b_{ij}=\hat{r}_i\hat{r}_j+\hat{s}_i\hat{s}_j, \qquad & e^l_{ij} =\hat{\Omega}_i\hat{\Omega}_j.
\end{split}
\end{equation}
In terms of the polarization tensor, the GW signal is $h_{ij}(t)=\sum_{A} e^A_{ij} h^A(t)$, where
$A=+,\times,x,y,b,l$ stands for the six  polarizations.

\subsection{The transfer function for one-way transmission}

\begin{figure}[htp]
\centering
	\includegraphics[width=0.4\textwidth]{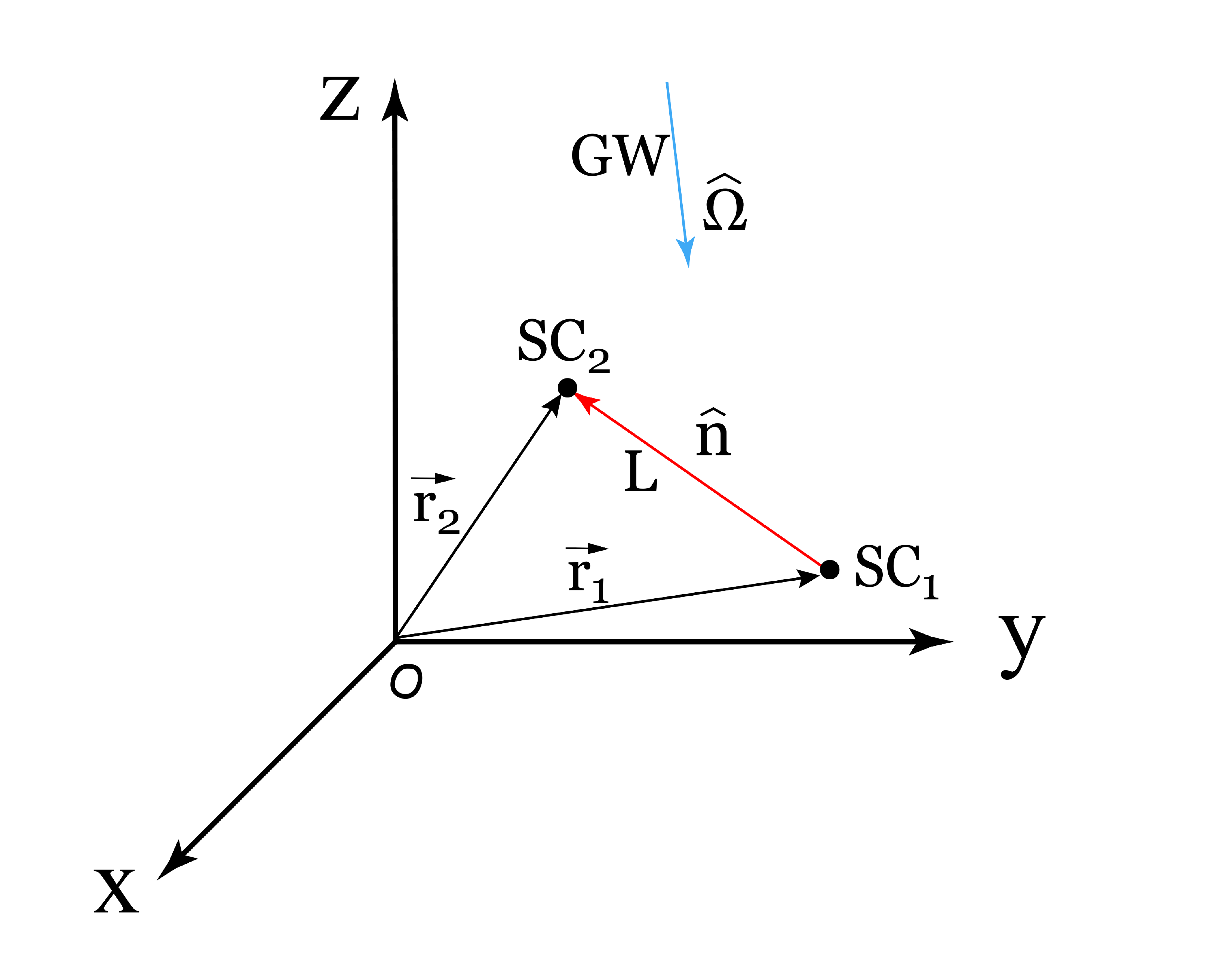}
\caption{One-way Doppler tracking. GWs propagate along $\hat{\Omega}$ and the laser beam transmits along $\hat{n}$ between two spacecrafts $SC_1$
located at $\vec{r}_1$ and $SC_2$ located at $\vec{r}_2$.
The arm length between $SC_1$ and $SC_2$ is $L$.}
\label{figgwtransf}
\end{figure}

Figure \ref{figgwtransf} shows the schematic geometry for one-way Doppler tracking.
The laser beams travel between $SC_1$ and $SC_2$ with the distance $L$ along the unit direction $\hat{n}$,
the one-way distance change due to GWs propagating along the unit direction $\hat{\Omega}$ is \cite{Larson:2002xr}
\begin{equation}
\label{deltal}
\begin{split}
	\delta L&=\sum_{A,i,j}\hat{n}_i\hat{n}_j e^A_{ij}\frac{\sin[\omega L(1-\hat{n}\cdot\hat{\Omega})/2]}{\omega(1-\hat{n}\cdot\hat{\Omega})}e^{-i\omega L(1-\hat{n}\cdot\hat{\Omega}+2\Omega\cdot\vec{r}_2/L)/2}h^A(f)\\
&=\sum_{A,i,j}L \hat{n}_i\hat{n}_j e^A_{ij}T(\omega,\hat{n}\cdot\hat{\Omega})h^A(f),
\end{split}
\end{equation}
where $\omega=2\pi f$ is the angular frequency of GWs. We take the speed of light $c=1$, and the transfer function $T(\omega,\hat{n}\cdot\hat{\Omega})$ for one-way Doppler tracking is \cite{Cornish:2001qi,Estabrook:1975}
\begin{equation}
\label{1transfeq}
\begin{split}
T(\omega,\hat{n}\cdot\hat{\Omega})&=\frac{\sin[\omega L(1-\hat{n}\cdot\hat{\Omega})/2]}{\omega L(1-\hat{n}\cdot\hat{\Omega})}e^{-i\omega L(1-\hat{n}\cdot\hat{\Omega}+2\Omega\cdot \vec{r}_2/L)/2}\\
&=\frac{1}{2}\text{sinc}\left[\omega L(1-\hat{n}\cdot\hat{\Omega})/2\right]e^{-i\omega L(1-\hat{n}\cdot\hat{\Omega}+2\Omega\cdot \vec{r}_2/L)/2},
\end{split}
\end{equation}
where $\text{sinc}(x)=\sin(x)/x$, $\vec{r}_2$ is the location of $SC_2$. In the low frequency limit, $\omega\rightarrow 0$, we have $T(\omega,\hat{n}\cdot\hat{\Omega})\to 1/2$.
For GWs with the propagation speed $v_{gw}$ different from the speed of light $c$, the transfer function is presented in Appendix \ref{appa}.
As discussed in Appendix \ref{appa}, the effect of propagation speed is negligible for space-based GW detectors, so we do not consider it for TDI.

\subsection{Response functions}

\begin{figure}[htp]
	\centering
	\includegraphics[width=0.4\textwidth]{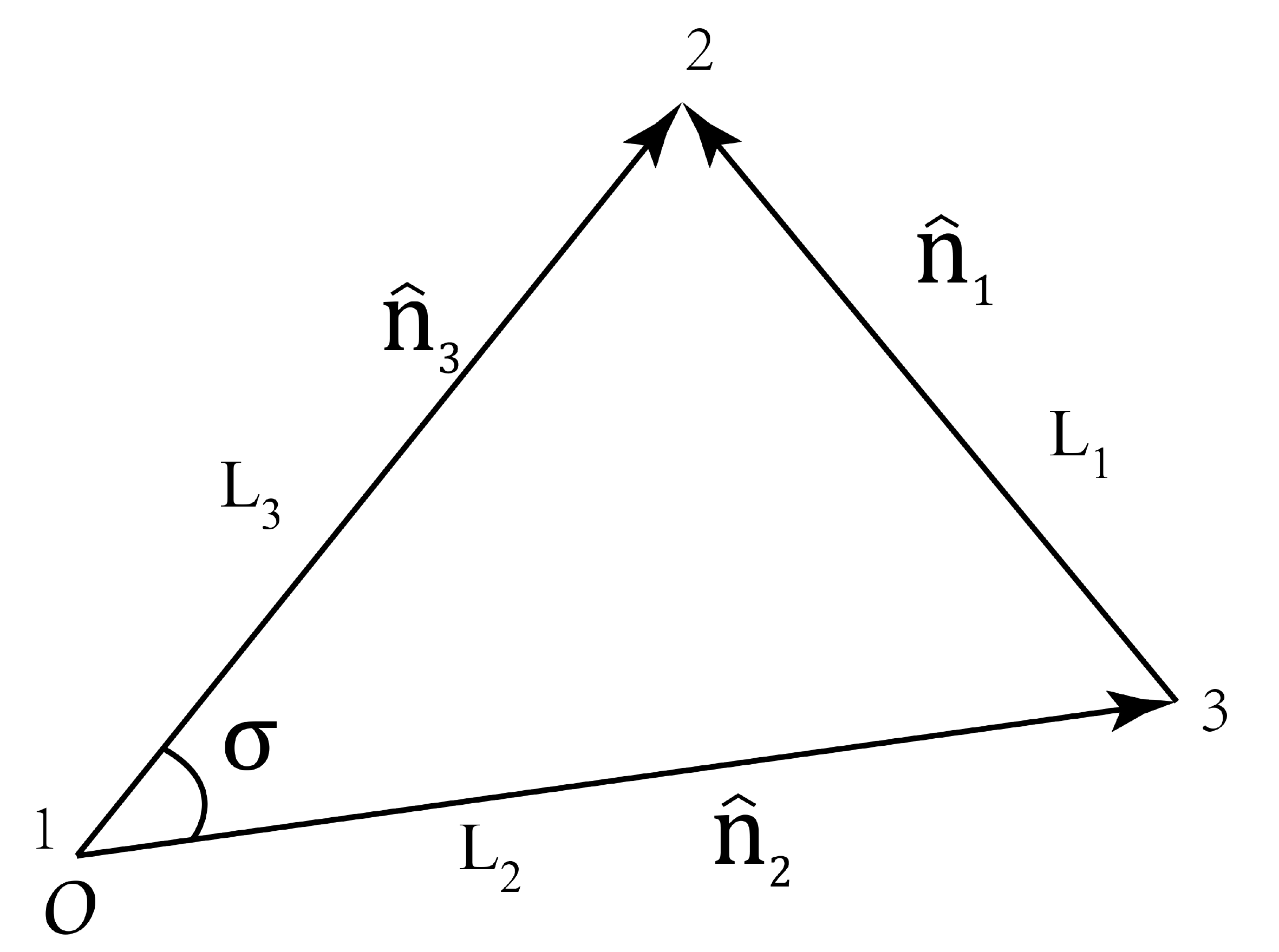}
\caption{Schematic configuration of a space-based, unequal arm interferometric GW detector. The spacecrafts are labeled as 1, 2 and 3,
and the spacecraft 1 ($SC_1$) is located at the origin. The optical paths between two SC pairs are denoted by $L_a$ and the unit vectors along the path are $\hat{n}_a$, with
the index $a$ corresponding to the opposite SC.}
\label{lisafig}
\end{figure}

\begin{figure}[htp]
	\centering
	\includegraphics[width=0.4\textwidth]{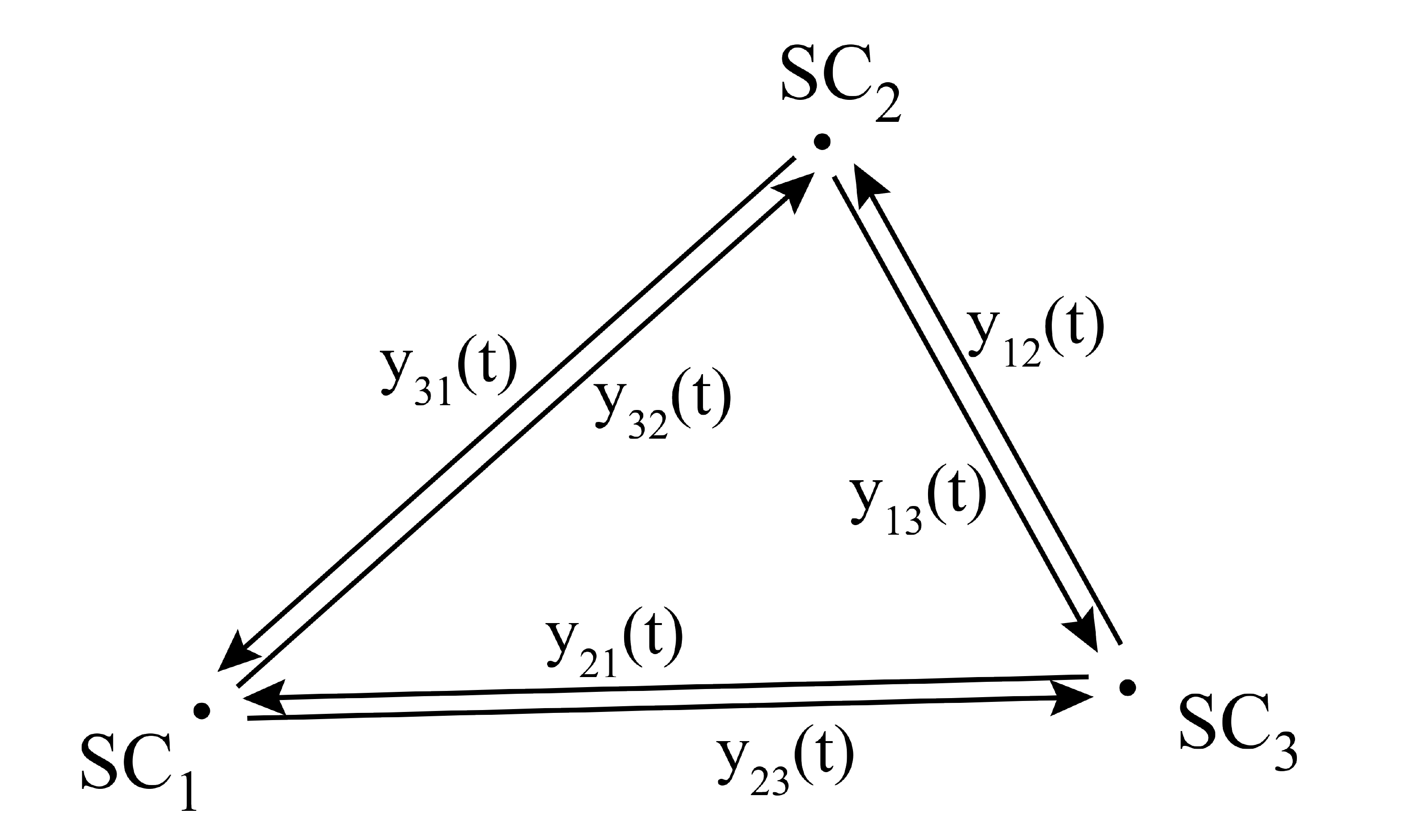}
	\caption{Six data beams $y_{ab}(t)$ exchanged between SCs.}
	\label{figdatabeams}
\end{figure}

Figure \ref{lisafig} shows the schematic configuration of a space-based, unequal arm interferometric GW detector.
The SC is labeled as 1, 2, 3, and $SC_1$ is located at the origin. The arm lengths between SC pairs
are $L_1$, $L_2$, $L_3$ and the unit vectors with the indicated orientation along the optical paths are $\hat{n}_a$, where $L_a$ and $\hat{n}_a$ are opposite to $SC_a$ and the index $a=1$, 2, 3 labels the SCs. Note that $L_1 \hat{n}_1+L_2 \hat{n}_2=L_3 \hat{n}_3$. Following \cite{Estabrook:2000ef,Prince:2002hp},
we denote $y_{ab}$ as the relative
frequency fluctuations time series measured from reception at $SC_b$ with transmission from $SC_d$ ($d\neq a$ and $d\neq b$) along $L_a$ as shown in Fig. \ref{figdatabeams}.
For example, $y_{31}$ is the relative frequency fluctuations time series measured from reception at $SC_1$ with transmission from $SC_2$ along $L_3$,
and the other five one-way relative frequency time series are obtained by cyclic permutation of the indices: $1\rightarrow 2\rightarrow 3\rightarrow 1$.
Similarly, the useful notation for delayed data streams are: $y_{31,2}=y_{31}(t-L_2)$, $y_{31,23}=y_{31}(t-L_2-L_3)=y_{31,32}$.

By using Eqs. \eqref{deltal} and \eqref{1transfeq}, we get
the GW response for the six TDI signal \cite{Estabrook:2000ef}
\begin{equation}
\label{yab}
y^{gw}_{ab}(T_D)=-\frac{\delta L_a}{L_e}=-\sum_{A,i,j}g_a\hat{n}^i_a\hat{n}^j_a e^A_{ij} T_{ab}(T_D)h^A(f),
\end{equation}
\begin{equation}
\label{t12eq}
T_{12}(T_D)=\frac{1}{2}\text{sinc}\left[\frac{1}{2} u_1(1-\mu_1)\right]e^{-i u_1(1-\mu_1)/2-i(\mu_3 u_3+T_D)},
\end{equation}
\begin{equation}
\label{t23eq}
T_{23}(T_D)=\frac{1}{2}\text{sinc}\left[\frac{1}{2} u_2(1-\mu_2)\right]e^{-i u_2(1-\mu_2)/2-i(\mu_2 u_2+T_D)},
\end{equation}
\begin{equation}
\label{t32eq}
T_{32}(T_D)=\frac{1}{2}\text{sinc}\left[\frac12 u_3(1-\mu_3)\right]e^{-i u_3(1-\mu_3)/2-i(\mu_3 u_3+T_D)},
\end{equation}
\begin{equation}
\label{t31eq}
T_{31}(T_D)=\frac{1}{2}\text{sinc}\left[\frac12 u_3(1+\mu_3)\right]e^{-i u_3(1+\mu_3)/2-iT_D},
\end{equation}
\begin{equation}
\label{t13eq}
T_{13}(T_D)=\frac{1}{2}\text{sinc}\left[\frac{1}{2} u_1(1+\mu_1)\right]e^{-i u_1(1+\mu_1)/2-i(\mu_2 u_2+T_D)},
\end{equation}
\begin{equation}
\label{t12eq}
T_{21}(T_D)=\frac{1}{2}\text{sinc}\left[\frac{1}{2} u_2(1+\mu_2)\right]e^{-i u_2(1+\mu_2)/2-iT_D},
\end{equation}
where $g_a=L_a/L_e$, $L_e$ is the expected arm length, $u=\omega L_e$,
$u_a=\omega L_a=g_a u $, $\mu_a=\hat{n}_a\cdot\hat{\Omega}$, and $T_D$ is the corresponding time delay. For example,
$T_D=u_1+u_2$ for $y^{gw}_{23,12}$. Note that to recover the 
results for $y_{ab}$ in \cite{Armstrong_1999,Estabrook:2000ef}, 
we need to multiply Eq. \eqref{yab} by $u$ because 
we use the fractional change $\delta L/L$.

The signal in the arm of the interferometric GW detector is
\begin{equation}
s(f)=\sum_A F^A h_A(f),
\end{equation}
where $F^A$ is the angular response function for polarization $A$. Averaging over source direction and polarization angle,
we get the averaged response function $R^A(f)$,
\begin{equation}
\label{ra}
	R^A(f)=\frac{1}{8\pi^2}\int_0^\pi\int_0^{2\pi}\int_0^{2\pi} |F^A(f,\theta,\phi,\psi)|^2\sin\theta d\theta d\phi d\psi.
\end{equation}

For the TDI $\alpha$ combination \cite{Armstrong_1999,Estabrook:2000ef},
the response functions $F^A_\alpha(f,\theta,\phi,\psi)$ are defined as
\begin{equation}
\label{ralphaeq}
\alpha^{gw}=y^{gw}_{21}-y^{gw}_{31}+y^{gw}_{13,2}-y^{gw}_{12,3}+y^{gw}_{32,12}-y^{gw}_{23,13}=\sum_A F^A_\alpha(f,\theta,\phi,\psi)h_A(f).
\end{equation}
Similarly, the response functions for other TDI combinations can be derived.

\section{analytical formulas of averaged response functions}
\label{sec3}

\begin{figure}[htp]
	\includegraphics[width=0.8\textwidth]{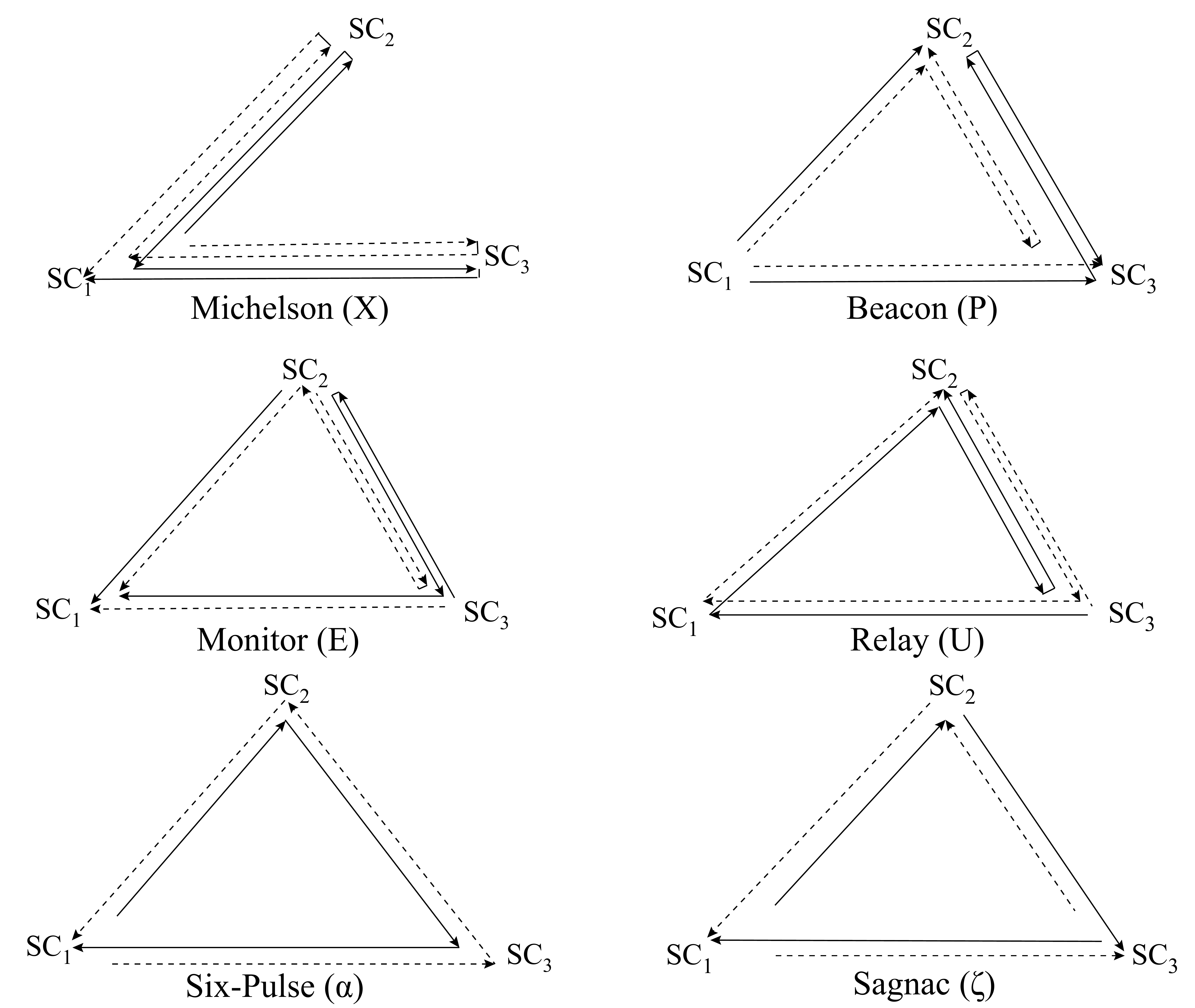}
\caption{Schematic diagrams for TDI $X$, $P$, $E$, $U$, 
$\alpha$, and $\zeta$ with the others in each class 
determined from cyclic permutation of indices.
Solid lines denote the path corresponding to the plus terms, and dashed lines denote the other path corresponding to the minus terms.}
\label{relayfig}
\end{figure}

There are six different TDI combinations: the six-pulse combination ($\alpha,\beta,\gamma$), the fully symmetric (Sagnac) combination ($\zeta$),
the unequal arm Michelson variables ($X,Y,Z$),
the beacon ($P,Q,R$), the monitor ($E,F,G$), and the relay ($U,V,W$) \cite{Armstrong_1999,Estabrook:2000ef}. we show them in Fig. \ref{relayfig}.
Combining Eqs. \eqref{yab}, \eqref{ra}, and \eqref{ralphaeq}, we can
derive the averaged response functions for all polarizations and all TDI combinations.
By assuming uniform distribution of the sources over the celestial sphere, the averaged response functions were plotted in \cite{Tinto:2010hz}
via Monte Carlo integration with 4000 source position/polarization state pairs per Fourier
frequency bin  and 7000 Fourier bins across the LISA band. In this section,
we derive analytical formulas for the averaged response functions and analyze their asymptotic behaviors. We work in the source
coordinate system and follow the notation used in \cite{Larson:1999we,Liang:2019pry}. In particular,
$\mu_1=\cos\theta_1=(\mu_2\cos\sigma+\sin\theta_2\sin\sigma\cos\epsilon)L_3/L_1-\mu_2 L_2/L_1$,
$\mu_3=\cos\theta_3=\mu_2\cos\sigma+\sin\theta_2\sin\sigma\cos\epsilon$, $\mu_2=\cos\theta_2$,
$\kappa_1=\sin\theta_2\cos\sigma-\mu_2\sin\sigma\cos\epsilon$, and
$\kappa_2=(\sin\theta_2\cos\sigma-\mu_2\sin\sigma\cos\epsilon)L_3/L_1-\sin\theta_2 L_2/L_1$.
$\epsilon$ is the angle between the plane containing $\hat{n}_2$ and $\hat{\Omega}$
and the plane of the interferometer,
$\sigma$ is the opening angle between two arms and $\cos\sigma=(L_2^2+L_3^2-L_1^2)/(2L_2L_3)$.
$\text{Si}(x)$ is the sine-integral function, and $\text{Ci(x)}$ is the cosine-integral function.
In the low frequency limit, $u=\omega L_e\to 0$,
\begin{equation}
\label{lowlim}
\begin{split}
{\rm Si}(2u)\to 2u-\frac{4u^3}{9}+\frac{4u^5}{75}+o(u^7),\\
{\rm Ci}(2u)\to\ln(2u)+\gamma_E-u^2+\frac{u^4}{6}-\frac{2u^6}{135}+o(u^8),
\end{split}
\end{equation}
where $\gamma_E$ is Euler number.
In the high frequency limit, $u\to\infty$,
\begin{equation}
\label{highlim}
\begin{split}
	{\rm Si}(u)\to\frac\pi2,\\
{\rm Ci}(u)\to 0.
\end{split}
\end{equation}

\subsection{Unequal arm Michelson combination}

The unequal arm Michelson variables $X$, $Y$, and $Z$ use only four beams and two laser beams exchanged between two of the SCs are not used.
The GW response for $X$ is $X^{gw}=y^{gw}_{32,322}-y^{gw}_{23,233}+y^{gw}_{31,22}-y^{gw}_{21,33}+y^{gw}_{23,2}-y^{gw}_{32,3}+y^{gw}_{21}-y^{gw}_{31}$.
The GW responses for $Y$ and $Z$ are obtained by the cyclic permutation of the indices,
 $Y^{gw}=y^{gw}_{13,133}-y^{gw}_{31,311}+y^{gw}_{12,33}-y^{gw}_{32,11}+y^{gw}_{31,3}-y^{gw}_{13,1}+y^{gw}_{32}-y^{gw}_{12}$,
and $Z^{gw}=y^{gw}_{21,211}-y^{gw}_{12,122}+y^{gw}_{23,11}-y^{gw}_{13,22}+y^{gw}_{12,1}-y^{gw}_{21,2}+y^{gw}_{13}-y^{gw}_{23}$.
The averaged response functions of $X$ are
\begin{equation}
\label{rxtensor}
\begin{split}
	u^2R_X^+(u)=&u^2R_X^\times(u)=
	2\sin^2u_3\left[\left(1+\cos^2u_2\right)\left(\frac13-\frac{2}{u_2^2}\right)
+\sin^2u_2+\frac{2\sin(2u_2)}{u^3_2}\right]\\
&+2\sin^2u_2\left[\left(1+\cos^2u_3\right)\left(\frac13-\frac{2}{u_3^2}\right)
+\sin^2u_3+\frac{2\sin(2u_3)}{u^3_3}\right]\\	
&-\frac1\pi\sin u_2 \sin u_3\int_0^{2\pi}d\epsilon\int_{-1}^1d\mu_2
\left(1-\frac{2\sin^2\sigma\sin^2\epsilon}{1-\mu_3^2}\right)\eta_X(\mu_2,\mu_3),
\end{split}
\end{equation}

\begin{equation}
\label{rxvector}
\begin{split}
	u^2R_X^x(u)=&u^2R_X^y(u)=
	4\sin^2u_3\left[2\gamma_E-5+2\ln (2u_2)-2{\rm Ci}(2u_2)-\frac13\cos(2u_2)+\right.\\
	&2\sin(2u_2)\left(\frac1{u_2}-\frac2{u_2^3}\right)+\left.\frac{4(1+\cos^2u_2)}{u_2^2}\right]+4\sin^2u_2\left[2\gamma_E-5+2\ln (2u_3)\right.\\
	&\left.-\frac13\cos(2u_3)-2{\rm Ci}(2u_3)+2\sin(2u_3)\left(\frac1{u_3}-\frac2{u_3^3}\right)+\frac{4(1+\cos^2u_3)}{u_3^2}\right]\\
	&-\frac4\pi\sin u_2\sin u_3
\int_0^{2\pi}d\epsilon\int_{-1}^1 d\mu_2\,\frac{\mu_2\mu_3\kappa_1}{(1-\mu_3^2)\sqrt{1-\mu_2^2}}\eta_X(\mu_2,\mu_3),
\end{split}
\end{equation}

\begin{equation}
\label{rxbeq}
\begin{split}
	u^2R_X^b(u)=&
	4\sin^2u_3\left[\left(1+\cos^2u_2\right)\left(\frac13-\frac{2}{u_2^2}\right)+\sin^2(u_2)
+\frac{2\sin(2u_2)}{u_2^3}\right]\\
	&+4\sin^2u_2\left[(1+\cos^2u_3)\left(\frac13-\frac{2}{u_3^2}\right)+\sin^2(u_3)+\frac{2\sin(2u_3)}{u_3^3}\right]\\
	&-\frac2{\pi}\sin u_2\sin u_3\int_0^{2\pi}d\epsilon \int_{-1}^1 d\mu_2\, \eta_X(\mu_2,\mu_3),
\end{split}
\end{equation}

\begin{equation}
\label{rxleq}
\begin{split}
	u^2R_X^l(u)=&\sin^2u_3\left[15-9\gamma_E-9\ln (2u_2)+\left(\frac{11}{3}-\gamma_E-\ln(2u_2)\right)\cos(2u_2)+\right.\\
	&\left(9+\cos(2u_2)\right){\rm Ci}(2u_2)+\left(2u_2+\sin(2u_2)\right){\rm Si}(2u_2)\\
&\left.+8\sin(2u_2)\left(\frac1{u_2^3}-\frac1{u_2}\right)-
\frac{8(1+\cos^2u_2)}{u_2^2}\right]\\
&+\sin^2u_2\left[15-9\gamma_E
-9\ln (2u_3)+\left(\frac{11}{3}-\gamma_E-\ln(2u_3)\right)\cos(2u_3)\right.\\
	&+\left(9+\cos(2u_3)\right){\text{Ci}}(2u_3)
	+\left(2u_3+\sin(2u_3)\right){\rm Si}(2u_3)\\
&\left.+8\sin(2u_3)\left(\frac1{u^3_3}-\frac1{u_3}\right)
	-\frac{8(1+\cos^2u_3)}{u_3^2}\right]\\
&-\frac2\pi\sin u_2\sin u_3\int_0^{2\pi} d\epsilon \int_{-1}^1 d\mu_2\,\frac{\mu_2^2\;\mu_3^2\; \eta_X(\mu_2,\mu_3)}{(1-\mu_2^2)(1-\mu_3^2)},
\end{split}
\end{equation}
where
\begin{equation}
\begin{split}
\eta_X(\mu_2,\mu_3)=\mu_2\mu_3[\cos u_2-\cos(u_2\mu_2)][\cos u_3-\cos(u_3\mu_3)]+\\
[\sin u_2-\mu_2\sin(u_2\mu_2)][\sin u_3-\mu_3\sin(u_3\mu_3)].
\end{split}
\end{equation}
Similarly, we can obtain the averaged response functions $R^A_Y$ and $R^A_Z$ by cyclic permutation of arm lengths ($u_1\to u_2\to u_3\to u_1$) from $R^A_X$. The integrals need to be calculated numerically.

\begin{figure}[htp]
\centering
	\includegraphics[width=0.8\textwidth]{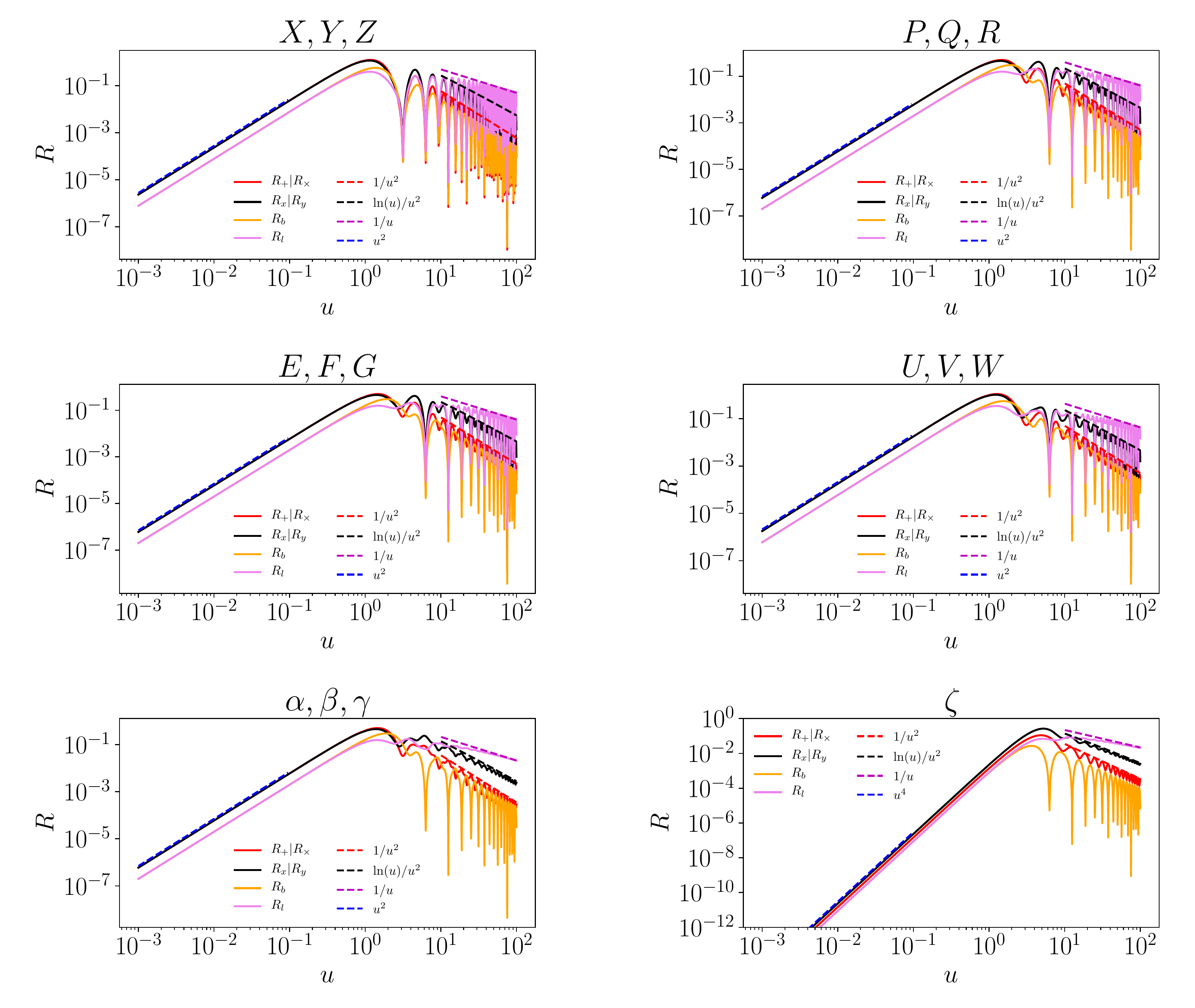}
\caption{Averaged response functions of different TDI combinations for tensor, vector, breathing and longitudinal modes. We assume equal arm lengths with $L_1=L_2=L_3=L_e$. After multiplying $u^2$ due to the definition we used in Eq. \eqref{yab}, these results are consistent with Fig. 3 in \cite{Tinto:2010hz}.}
\label{rufig}
\end{figure}

From Eqs. \eqref{rxtensor}, \eqref{rxvector}, \eqref{rxbeq}, and \eqref{rxleq}, it is easy to show that in the low frequency limit, $\omega\to 0$,
$R_X^A \propto \omega^2$. In the high frequency limit, we find that $R_X \propto 1/\omega^2$ for the tensor and breathing modes,
$R_X \propto 1/\omega$ for the longitudinal mode, and $R_X \propto \ln(\omega)/\omega^2$ for the vector mode.
The high frequency behaviors are the same as those for the equal arm interferometric
GW detector without optical cavities derived in \cite{Liang:2019pry}.
We show the averaged response functions of the TDI combinations $X$, $Y$, and $Z$
for space-based interferometers with equal arm lengths in Fig. \ref{rufig}.
As shown in Fig. \ref{rufig}, at frequencies equal to half-integer multiple of $1/L_e$, the averaged response functions for all polarizations are zero
due to the overall factor $\sin^2 u$ in Eqs. \eqref{rxtensor}, \eqref{rxvector}, \eqref{rxbeq}, and \eqref{rxleq},
so the response of the detector is subtracted to zero at those frequencies.
Since a tensor signal propagating perpendicularly to the light beam interacts with the light during the brief instance its wave front
crosses the light beam, and a scalar longitudinal wave propagating along the light beam affects the light during the time
it travels the entire arm length $L$, so the averaged response to the scalar longitudinal wave is significantly larger than
that to the tensor signal when the wavelength of GWs is shorter than the arm length \cite{Tinto:2010hz}.

\subsection{Beacon combinations $P$, $Q$, $R$}

For the TDI Beacon combinations $P$, $Q$ and $R$, one SC emits a laser only and does not receive the laser from the other two SCs.
The GW response for $P$ is $P^{gw}=y^{gw}_{32,2}-y^{gw}_{23,3}-y^{gw}_{12,2}+y^{gw}_{13,3}+y^{gw}_{12,13}-y^{gw}_{13,12}+y^{gw}_{23,311}-y^{gw}_{32,211}$.
The GW responses for $Q$ and $R$ are obtained by the cyclic permutation of the indices. The analytical expressions for
the averaged response functions of $P$ are presented in Appendix \ref{beacon}.
From Eqs. \eqref{rptensor}, \eqref{rpvector}, \eqref{rpbeq} and \eqref{rpleq},
it is easy to see that in the low frequency limit,
$R_P^A \propto \omega^2$. In the high frequency limit,
we find that $R_P \propto 1/\omega^2$ for the tensor and breathing modes,
$R_P \propto 1/\omega$ for the longitudinal mode and $R_P \propto \ln(\omega)/\omega^2$ for the vector mode,
which are the same as those for the equal arm interferometric
GW detector without optical cavities derived in \cite{Liang:2019pry}.
We show the averaged response functions of the TDI Beacon combinations $P$, $Q$ and $R$
for space-based interferometers with equal arm lengths in Fig. \ref{rufig}.
From Fig. \ref{rufig}, we see that the averaged response functions for all polarizations are very small at frequencies equal to integer multiples of $1/L_e$
due to the factor $1-\cos u$ and $\sin^2 u$ in Eqs. \eqref{rptensor}, \eqref{rpvector}, \eqref{rpbeq}, and \eqref{rpleq}.

\subsection{Monitor combinations}

For the TDI monitor combinations $E$, $F$, and $G$, one spacecraft only receives laser and will not emit laser beams to the other two spacecrafts.
The GW response for $E$ is $E^{gw}=y^{gw}_{12,21}-y^{gw}_{13,31}-y^{gw}_{12,3}+y^{gw}_{13,2}+y^{gw}_{31,11}-y^{gw}_{21,11}-y^{gw}_{31}+y^{gw}_{21}$.
The GW responses for $F$ and $G$ are obtained by the cyclic permutation of the indices. The analytical expressions for
the averaged response functions of $E$ are presented in Appendix \ref{monitor}.
From Eqs. \eqref{retensor}, \eqref{revector}, \eqref{rebeq}, and \eqref{releq},
it is easy to see that in the low frequency limit,
$R_E^A \propto \omega^2$. In the high frequency limit, we find that $R_E \propto 1/\omega^2$ for the tensor and breathing modes,
$R_E \propto 1/\omega$ for the longitudinal mode and $R_E \propto \ln(\omega)/\omega^2$ for the vector mode,
which are the same as those for the equal arm interferometric
GW detector without optical cavities derived in \cite{Liang:2019pry}.
We show the averaged response functions of the TDI monitor combinations $E$, $F$ and $G$
for space-based interferometers with equal arm lengths in Fig. \ref{rufig}.
From Fig. \ref{rufig}, we see that the averaged response functions for all polarizations are very small at frequencies equal to integer multiples of $1/L_e$
due to the factor $1-\cos u$ and $\sin^2 u$ in Eqs. \eqref{retensor}, \eqref{revector}, \eqref{rebeq}, and \eqref{releq}.

\subsection{Relay combinations}

For the TDI relay combinations $U$, $V$, and $W$, one spacecraft receives laser beam from another one, and emits a laser beam to the remaining one. The GW response for $U$ is
$U^{gw}=y^{gw}_{21,113}-y^{gw}_{21,3}-y^{gw}_{12,123}+y^{gw}_{13,1}-y^{gw}_{13,23}+y^{gw}_{32,11}-y^{gw}_{32}+y^{gw}_{12}$.
The GW responses for $V$ and $W$ are obtained by the cyclic permutation of the indices. The analytical expressions for
the averaged response functions of $U$ are presented in Appendix \ref{relay}.
From Eqs. \eqref{rutensor}, \eqref{ruvector}, \eqref{rubeq}, and \eqref{ruleq},
it is easy to see that in the low frequency limit,
$R_U^A \propto \omega^2$. In the high frequency limit, we find that $R_U \propto 1/\omega^2$ for the tensor and breathing modes,
$R_U \propto 1/\omega$ for the longitudinal mode and $R_U \propto \ln(\omega)/\omega^2$ for the vector mode,
which are the same as those for the equal arm interferometric
GW detector without optical cavities derived in \cite{Liang:2019pry}.
We show the averaged response functions of the TDI monitor combinations $U$, $V$ and $W$
for space-based interferometers with equal arm lengths in Fig. \ref{rufig}.
From Fig. \ref{rufig}, we see that the averaged response functions for all polarizations are very small at frequencies equal to integer multiple of $1/L_e$
due to the factor $1-\cos u\cos 2u$ and $\sin^2 u$ in Eqs. \eqref{rutensor}, \eqref{ruvector}, \eqref{rubeq}, and \eqref{ruleq}.

\subsection{The $\alpha$, $\beta$, $\gamma$ (six-pulse) combinations}

For the TDI combinations $\alpha$, $\beta$, and $\gamma$, all six laser beams are used.
The GW response for $\alpha$ is $\alpha^{gw}=y^{gw}_{21}-y^{gw}_{31}+y^{gw}_{13,2}-y^{gw}_{12,3}+y^{gw}_{32,12}-y^{gw}_{23,13}$.
The GW responses for $\beta$ and $\gamma$ are obtained by the cyclic permutation of the indices. The analytical expressions for
the averaged response functions of $\alpha$ are presented in Appendix \ref{alpha}.
From Eqs. \eqref{ratensor}, \eqref{ravector}, \eqref{rabeq}, and \eqref{raleq},
it is easy to see that in the low frequency limit,
$R_\alpha^A \propto \omega^2$. In the high frequency limit, we find that $R_\alpha \propto 1/\omega^2$ for the tensor and breathing modes,
$R_\alpha \propto 1/\omega$ for the longitudinal mode and $R_\alpha \propto \ln(\omega)/\omega^2$ for the vector mode,
which are the same as those for the equal arm interferometric
GW detector without optical cavities derived in \cite{Liang:2019pry}.
We show the averaged response functions of the TDI combinations $\alpha$, $\beta$,
and $\gamma$ for space-based interferometers with equal arm lengths in Fig. \ref{rufig}.
From Fig. \ref{rufig}, we see that the averaged response function for breathing mode is almost zero at frequencies equal to integer multiples of $1/L_e$ because the integral term almost cancels the constant 4+4/3 in Eq. \eqref{rabeq}.

\subsection{Fully symmetric (Sagnac) combination}

The fully symmetric (Sagnac) TDI combination $\zeta$ uses all six laser beams. The GW response for $\zeta$ is $\zeta^{gw}=y^{gw}_{32,2}-y^{gw}_{23,3}+y^{gw}_{13,3}-y^{gw}_{31,1}+y^{gw}_{21,1}-y^{gw}_{12,2}$.
The analytical expressions for
the averaged response functions of $\zeta$ are presented in appendix \ref{sagnac}.
From Eqs. \eqref{rztensor}, \eqref{rzvector}, \eqref{rzbeq} and \eqref{rzleq},
it is easy to see that in the low frequency limit,
$R_\zeta^A \propto \omega^2$. However, for the equilateral-triangle case $L_1=L_2=L_3=L_e$, $R_\zeta^A \propto \omega^4$.
In the high frequency limit, we find that $R_\zeta \propto 1/\omega^2$ for the tensor and breathing modes,
$R_\zeta \propto 1/\omega$ for the longitudinal mode and $R_\zeta \propto \ln(\omega)/\omega^2$ for the vector mode,
which are the same as those for the equal arm interferometric
GW detector without optical cavities derived in \cite{Liang:2019pry}.
We show the averaged response functions of the TDI Sagnac combination $\zeta$ for
space-based interferometers with equal arm lengths in Fig. \ref{rufig}.
From Fig. \ref{rufig}, we see that the averaged response function for breathing mode is very small at frequencies
equal to integer multiple of $1/L_e$ because the integral term almost cancels the constant 4+4/3 in Eq. \eqref{rzbeq}.

\section{sensitivity curves}
\label{sec4}

Armed with the averaged response functions for different TDI combinations, we are ready to plot the sensitivity curves.
For LISA, the single-link optical metrology noise is \cite{Cornish:2018dyw}
\begin{equation}
\label{lisasx}
P_{\text{OMS}}=(1.5\times10^{-11}{\rm m})^2\left[1+\left(\frac{\rm 2mHz}{f}\right)^4\right]{\text{Hz}}^{-1},
\end{equation}
and the acceleration noise is
\begin{equation}
\label{lisasa}
P_a=(3\times10^{-15}{\rm m\ s^{-2}})^2\left[1+\left(\frac{0.4{\rm mHz}}{f}\right)^2\right]\left[1+\left(\frac{f}{\rm 8mHz}\right)^4\right]{\rm\ Hz^{-1}}.
\end{equation}
Dividing them by $L_e^2$ and $L_e^2(2\pi f)^4$ respectively, we get the shot and proof mass noises
\begin{equation}
\label{lisan}
\begin{split}
	S^{\text{shot}}_y=&3.6\times10^{-41}\left[1+\left(\frac{\rm 2mHz}{f}\right)^4\right]\ \text{Hz}^{-1},\\
	S^{\text{proof mass}}_y=&9.24\times10^{-52}\left(\frac{\text{Hz}}{f}\right)^4 \left[1+\left(\frac{0.4{\rm mHz}}{f}\right)^2\right]\left[1+\left(\frac{f}{\rm 8mHz}\right)^4\right] \ \text{Hz}^{-1}.
\end{split}
\end{equation}

For TianQin, the position and acceleration noises are $S_x=10^{-24}{\rm m^2/Hz}$ and $S_a=10^{-30}{\rm m^2s^{-4}/Hz}$ \cite{Luo:2015ght}.
So the shot and proof mass noises are
\begin{equation}
\label{tqn}
\begin{split}
S^{\text{shot}}_y=&3.33 \times 10^{-41} \ \text{Hz}^{-1},\\
S^{\text{proof mass}}_y=&2.14 \times10^{-50}\left(\frac{\text{Hz}}{f}\right)^2 \ \text{Hz}^{-1}.
\end{split}
\end{equation}

\begin{figure}[htp]
\centering
	\includegraphics[width=0.8\textwidth]{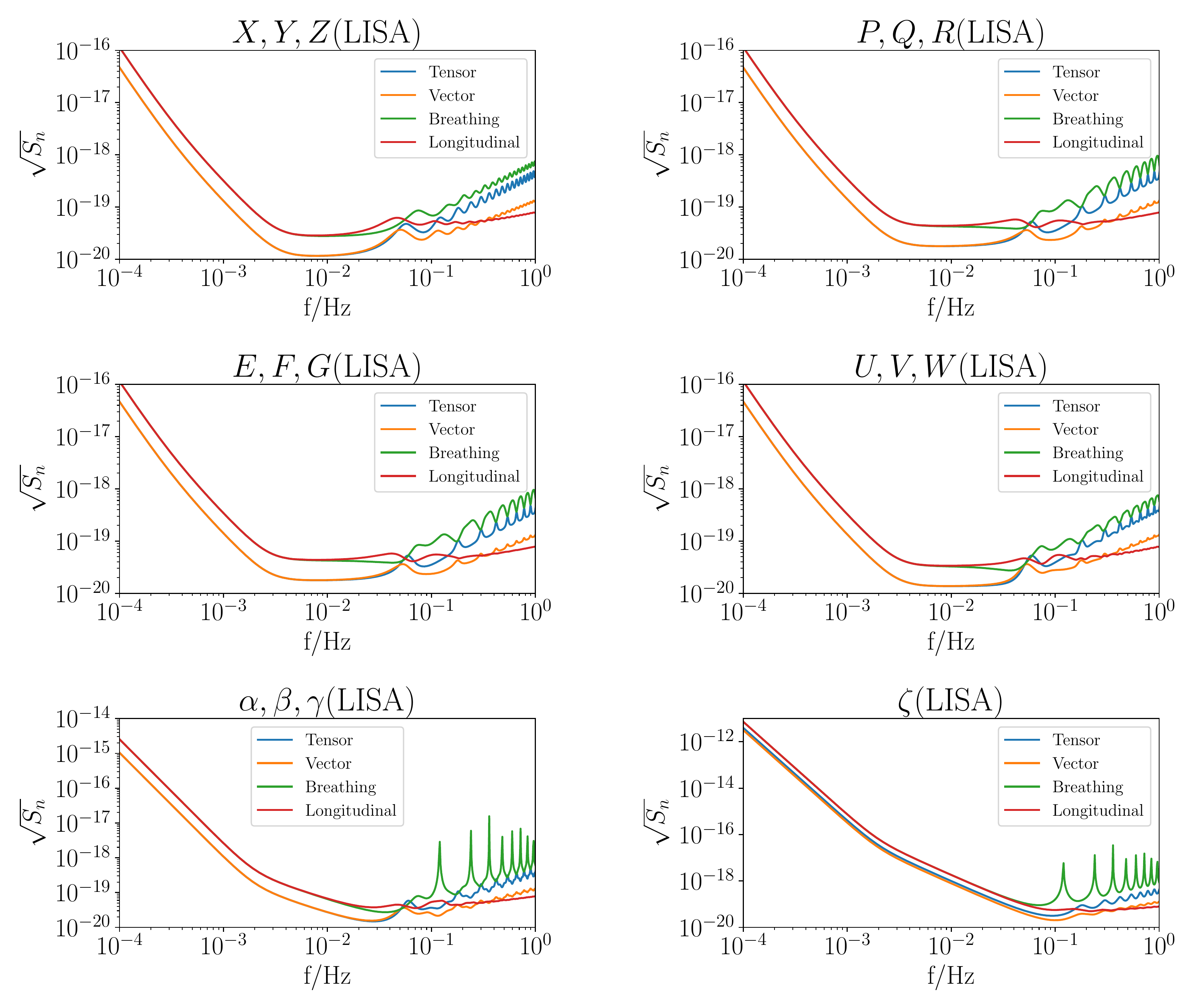}
\caption{LISA sensitivities of different TDI combinations for tensor, vector, breathing and longitudinal polarizations. We take $L_1=L_2=L_3=2.5\times10^9$ m.}
\label{lisascurve}
\end{figure}

\begin{figure}[htp]
\centering
	\includegraphics[width=0.8\textwidth]{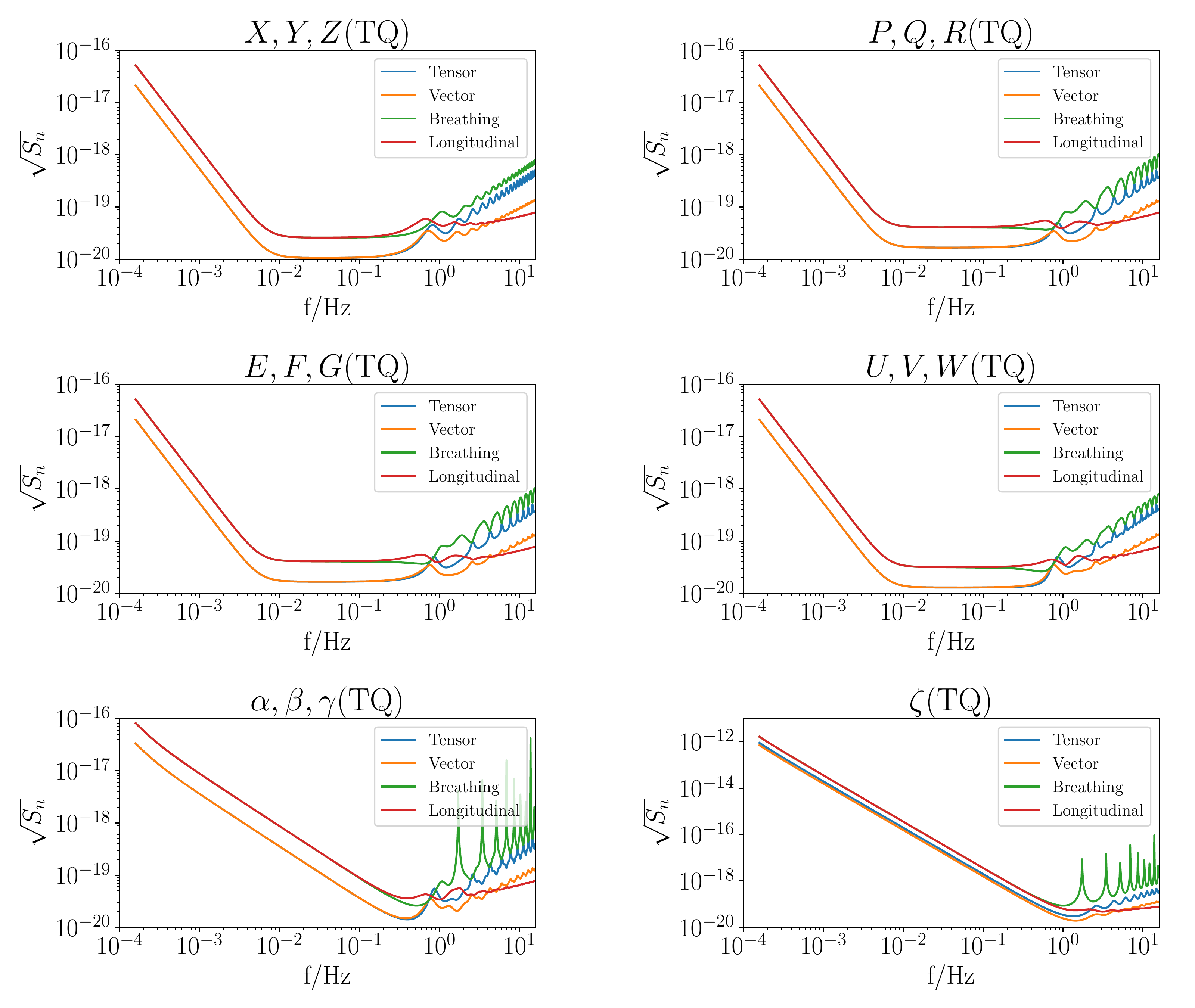}
\caption{TianQin sensitivities of different TDI combinations for tensor, vector, breathing, and longitudinal polarizations. We take $L_1=L_2=L_3=1.73\times 10^8$ m.}
\label{tqscurve}
\end{figure}

The noises in TDI combinations are \cite{Estabrook:2000ef}
\begin{equation}
\label{pnalpha}
P_n^\alpha=[8\sin^2(3\pi f L)+16\sin^2(\pi fL)]S_y^{\rm proof\ mass}+6 S_y^{\rm shot}, \end{equation}

\begin{equation}
\label{pnzeta}
P_n^\zeta=24\sin^2(\pi fL) S_y^{\rm proof\ mass}+6S_y^{\rm shot},
\end{equation}

\begin{equation}
\label{pnx}
P_n^X=[8\sin^2(4\pi f L)+32\sin^2(2\pi fL)]S_y^{\rm proof\ mass}+16\sin^2(2\pi fL)S_y^{\rm shot},
\end{equation}

\begin{equation}
\label{pnp}
P_n^P=[8\sin^2(2\pi f L)+32\sin^2(\pi fL)]S_y^{\rm proof\ mass}+[8\sin^2(2\pi fL)+8\sin^2(\pi fL)]S_y^{\rm shot},
\end{equation}

\begin{equation}
\label{pne}
P_n^E=[32\sin^2(\pi f L)+8\sin^2(2\pi fL)]S_y^{\rm proof\ mass}+[8\sin^2(\pi fL)+8\sin^2(2\pi fL)]S_y^{\rm shot},
\end{equation}

\begin{equation}
\label{pnu}
\begin{split}
P_n^U=&[16\sin^2(\pi f L)+8\sin^2(2\pi fL)+16\sin^2(3\pi fL)]S_y^{\rm proof\ mass}\\
&+[4\sin^2(\pi fL)+8\sin^2(2\pi fL)+4\sin^2(3\pi fL)]S_y^{\rm shot},
\end{split}
\end{equation}
The sensitivity curve for the tensor mode is
\begin{equation}
\label{tenscurve}
S_n^k=\frac{P_n^k}{R_k^++R_k^\times}=\frac{1}{2}\frac{P_n^k}{R_k^+},
\end{equation}
where $k=\alpha,\zeta,X,P,E,U$.
The sensitivity curve for the vector mode is
\begin{equation}
\label{vecscurve}
S_n^k=\frac{P_n^k}{R_k^x+R_k^y}=\frac{1}{2}\frac{P_n^k}{R_k^x}.
\end{equation}
The sensitivity curve for the breathing mode is
\begin{equation}
\label{bscurve}
S_n^k=\frac{P_n^k}{R_k^b}.
\end{equation}
The sensitivity curve for the longitudinal mode is
\begin{equation}
\label{longscurve}
S_n^k=\frac{P_n^k}{R_k^l}.
\end{equation}
Figure \ref{lisascurve} shows LISA sensitivities of different TDI combinations and Fig. \ref{tqscurve} shows TianQin sensitivities of different TDI combinations.
As noted in \cite{Tinto:2010hz}, there is a lack of sensitivity of the TDI combinations $\alpha$ and $\zeta$ to the breathing mode at frequencies equal to
integer multiples of $1/L_e$. However, for the unequal arm Michelson, beacon, monitor and relay TDI combinations,
at frequencies where the signal is lost the noises also cancel out, so there is no problem of lack of sensitivity at those frequencies.
In Ref. \cite{Tinto:2010hz}, the sensitivity is defined as the GW amplitude required
to achieve a signal-to-noise ratio 5 in a one-year integration time, so the amplitude
is about $5\sqrt{3.17\times 10^{-8}}=8.9\times 10^{-4}$ smaller. Taking into account this factor and the difference in the assumption about noise levels and arm lengths, the results in Fig. \ref{lisascurve} are consistent with those shown in Fig. 4 in Ref. \cite{Tinto:2010hz}.

\section{summary and conclusions}
\label{sec5}

We obtain analytical formulas for the averaged response functions of massive gravitons for tensor, vector, breathing and longitudinal polarizations.
For space-based GW detectors, the frequency is much larger than the mass of gravitons constrained by the observations, so the propagation speed of GWs does not differ much
from the speed of light and its effect on the averaged response functions is negligible. On the other hand, because the distance between SCs is comparable or even larger than the wavelength of in-band
GWs and it is impossible to keep the precise equality of the arm lengths, the TDI technique is needed to reduce the laser frequency noise by synthesizing virtual equal arm interferometric measurements.
We give analytical formulas for the averaged response functions of different TDI combinations for tensor, vector, breathing, and longitudinal modes
which are consistent with those obtained by numerical simulation.
With the analytical expressions, we obtain the asymptotic behaviors of the transfer functions.
At low frequencies, $f\ll f_*=1/(2\pi L_e)$,  the averaged response functions of all TDI combinations increase as $f^2$ for all six polarizations.
The one exception is that the averaged response functions of $\zeta$ for all six polarizations increase as $f^4$ in the equilateral-triangle case.
At high frequencies, $f\gg f_*$, the averaged response functions of all TDI combinations for the tensor and breathing modes fall off as $1/f^2$,
the averaged response functions of all TDI combinations for the vector mode fall off as $\ln(f)/f^2$ ,
and the averaged response functions of all TDI combinations for the longitudinal mode fall as $1/f$.
By using the analytical expressions for the averaged response functions of all TDI combinations,
we plot the frequency dependent response functions for tensor, vector, breathing and longitudinal polarizations with equal arms,
we also plot the LISA and TianQin sensitivity curves with different TDI combinations for tensor, vector, breathing and longitudinal polarizations.

\section*{acknowledgments}
This research was supported in part by the National Natural Science
Foundation of China under Grants No. 11875136 and 11605061,
the Major Program of the National Natural Science Foundation of China under Grant No. 11690021.
A.J.W. acknowledges funding from the U.S. National Science Foundation, under Cooperative Agreement No. PHY-1764464.
Q.G. acknowledges the Fundamental Research Funds for the Central Universities under Grants No. XDJK2017C059 and No. SWU116053,
the financial support from China Scholarship Council for sponsoring her visit to California Institute of Technology,
and the hospitality of California Institute of Technology.

\appendix
\section*{The transfer function for massive gravitons}
\label{appa}

For GWs with the propagation speed $v_{gw}$ which is different from the speed of light $c$,
the transfer function $T(f,\hat{n}\cdot\hat{\Omega})$ for a single round trip in the arm is \cite{Tinto:2010hz,Blaut:2015qaa}
\begin{equation}
\label{transferfunction}
\begin{split}
T(f,\hat{n}\cdot\hat{\Omega})=\frac{1}{2}& \left\{ \text{sinc}\left[\frac{f}{2f^\star}(1-\hat{n}\cdot\hat{\Omega}/(v_{gw}/c))\right]\exp\left[-i\frac{f}{2f^\star}(3+\hat{n}\cdot\hat{\Omega}/(v_{gw}/c))\right] \right.\\
&\left. +\text{sinc}\left[\frac{f}{2f^\star}(1+\hat{n}\cdot\hat{\Omega}/(v_{gw}/c))\right]\exp\left[-i\frac{f}{2f^\star}(1+\hat{n}\cdot\hat{\Omega}/(v_{gw}/c))\right]\right\},
\end{split}
\end{equation}
where $f^\star=c/(2\pi L)$ is the transfer frequency, $L$ is the arm length of the detector and here we write out $c$ explicitly.
For massive gravitons, the propagation speed for GWs is $v_{gw}(\omega)=c\sqrt{1-m_g^2/\omega^2}$.

\begin{figure}[htp]
\centering
	\includegraphics[width=0.8\textwidth]{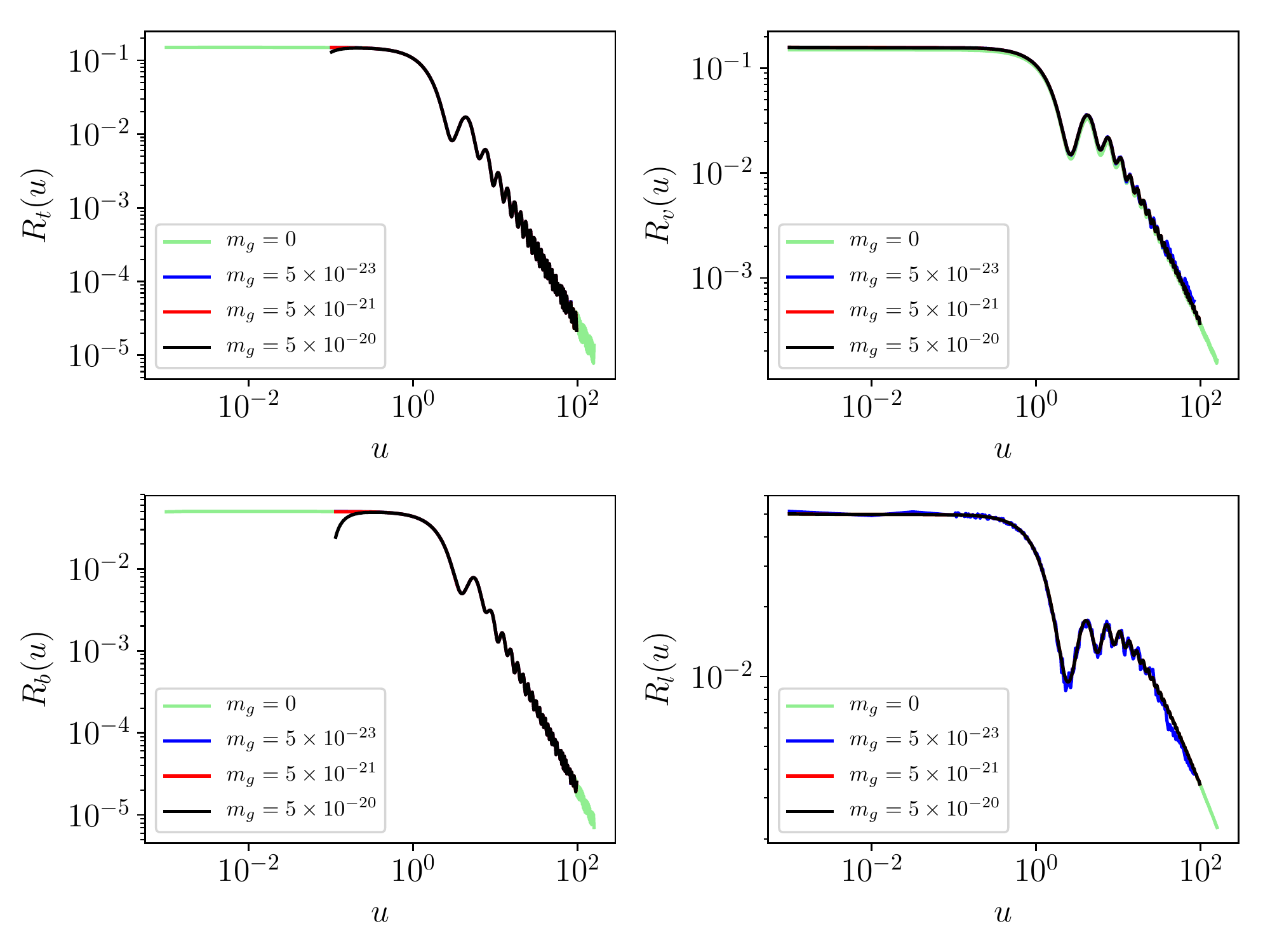}
\caption{Averaged response functions of massive gravitons for tensor, vector, breathing, and longitudinal polarizations.
We consider equal arm interferometric GW detectors ($L=1.73\times 10^8$ m for TianQin is used) without optical cavities in the arms.}
\label{rufige}
\end{figure}

Choosing the source coordinate system, we calculate the averaged angular response function for equal arm interferometric GW detectors without optical cavities in the arms,
and the results are
\begin{equation}
\begin{split}
R^{+}(u)=R^{\times}(u)&=H(u)-\frac{1}{16\pi u^{2}}\int_{0}^{2\pi}d\epsilon\int_{0}^{\pi}d\theta_{1}\sin^3(\theta_1)\\
&\qquad\qquad \times \left[\sin^2(\theta_2)-2\sin^2(\sigma)
\sin^2(\epsilon)\right]\eta(u),
\end{split}
\end{equation}

\begin{equation}
R^b(u)=2H(u)-\frac{1}{8\pi u^{2}}\int_{0}^{2\pi}d\epsilon\int_{0}^{\pi}d\theta_{1}\sin^3(\theta_{1})\sin^2(\theta_{2})\eta(u),
\end{equation}

\begin{equation}
\begin{split}
R^{x}(u)=&R^{y}(u)\\
=&\frac{v_{gw}^2}{12 u^5}\left\{
12 v_{gw}^2 \cos (u) \left[4 v_{gw} \left(u^2-1\right) \sin \left(\frac{u}{v_{gw}}\right)+u \left(u^2+4\right) \cos \left(\frac{u}{v_{gw}}\right)\right]\right.\\
&+2\left(2-3 v_{gw}^2\right) u^3 \cos(2u)+6 \left(2-9 v_{gw}^2\right) u^3\\
&\left.+2u\left[6 v_{gw}^3 \left(u^2+2\right) \sin (u) \sin \left(\frac{u}{v_{gw}}\right)-12 v_{gw}^2 u \sin (u) \cos \left(\frac{u}{v_{gw}}\right)\right]\right\}\\
&+\frac{v_{gw}^3}{2 u^2}\left(1-v_{gw}^2\right) \left[u+\sin (u) \cos (u)\right]\left[\text{Si}\left(u+\frac{u}{v_{gw}}\right)-\text{Si}\left(u-\frac{u}{v_{gw}}\right)\right]\\
&+\frac{v_{gw}^3}{4 u^2}\left[5-9v_{gw}^2+\left(1-v_{gw}^2\right)
\cos (2 u)\right]\left[\ln \left(\frac{1-v_{gw}}{1+v_{gw}}\right)\right.\\
&\left.\qquad \qquad +\text{Ci}\left(u+\frac{u}{v_{gw}}\right)-
\text{Ci}\left(\frac{u}{v_{gw}}-u\right)\right]\\
&-\frac{1}{8\pi u^{2}}\int_{0}^{2\pi}d\epsilon\int_{0}^{\pi}d\theta_1 \sin(\theta_{1})\sin(2\theta_{1})\cos(\theta_{2})\\
&\qquad\qquad\qquad \times \left[\cos(\sigma)\sin(\theta_{1})-\sin(\sigma)\cos(\theta_{1})\cos(\epsilon)\right]\eta(u),
\end{split}
\end{equation}

\begin{equation}
\begin{aligned}
R^l(u)=&\frac{v_{gw}^4}{u^3}\sin(u)\cos\left(\frac{u}{v_{gw}}\right)
-\frac{v_{gw}^5}{u^4}\sin(u)\sin\left(\frac{u}{v_{gw}}\right)
-\frac{2v_{gw}^4}{u^4}\cos(u)\cos\left(\frac{u}{v_{gw}}\right)\\
&+\frac{v_{gw}^2}{12u^2}(3v_{gw}^2+1)\cos(2u)-\frac{v_{gw}^3}{u^5}(2u^2v_{gw}^2-2v_{gw}^2+u^2)\cos(u)\sin\left(\frac{u}{v_{gw}}\right)\\
&+\frac{v_{gw}^2}{4(1-v_{gw}^2)u^2}\left[1-9v_{gw}^4+6v_{gw}^2\right.\\
&\qquad\qquad\qquad \left.+2v_{gw}^5\sin(u)\sin(\frac{u}{v_{gw}})+2v_{gw}^4\cos(u)\cos(\frac{u}{v_{gw}})\right]\\
&+\frac{\cos (2 u)+9}{8 u^2}v_{gw}^5 \left[\text{Ci} \left(u+\frac{u}{v_{gw}}\right)-\text{Ci} \left(\frac{u}{v_{gw}}-u\right)+\ln \frac{1-v_{gw}}{1+v_{gw}}\right]\\
&+\frac{ u+\sin (u)\cos(u)}{4u^2}v_{gw}^5\left[\text{Si}\left(\frac{u}{v_{gw}}+u\right)
-\text{Si}\left(u-\frac{u}{v_{gw}}\right)\right]\\
&-\frac{1}{8\pi u^{2}}\int_{0}^{2\pi}d\epsilon\int_{0}^{\pi}
d\theta_{1}\sin(\theta_{1})\cos^2(\theta_{1})\cos^2(\theta_{2})\eta(u),
\end{aligned}
\end{equation}
where
\begin{equation}
\begin{split}
H(u)=&\frac{-v_{gw}^2}{24 u^5}\left\{3 u^3 \left(7-9 v_{gw}^2\right)
+u^3 \left(5-3 v_{gw}^2\right) \cos(2u)
-12 u^2 v_{gw}^2 \sin (u) \cos \left(\frac{u}{v_{gw}}\right)
\right.\\
&+6u v_{gw} \sin (u) \left[\left(u^2+2\right) v_{gw}^2-u^2\right] \sin \left(\frac{u}{v_{gw}}\right)\\
&+6 \cos (u)\left(4v_{gw}^3 u^2-4v_{gw}^3 -2 v_{gw} u^2\right) \sin \left(\frac{u}{v_{gw}}\right)\\
&\left.+6 \cos (u)(u^3v_{gw}^2+4u v_{gw}^2-u^3) \cos \left(\frac{u}{v_{gw}}\right)\right\} \\
&+\frac{v_{gw}}{16 u^2} \left(v_{gw}^2-1\right) \left[ \left(v_{gw}^2-1\right) \cos (2 u)+9 v_{gw}^2-1\right]\\
&\qquad \qquad \times
\left[\text{Ci} \left(u+\frac{u}{v_{gw}}\right)-\text{Ci} \left(\frac{u}{v_{gw}}-u\right)+\ln \left(\frac{1-v_{gw}}{1+v_{gw}}\right)\right]\\
&+\frac{2 u+\sin (2 u)}{16 u^2}v_{gw} \left(v_{gw}^2-1\right)^2\left[\text{Si}\left(\frac{u}{v_{gw}}+ u\right)-\text{Si}\left(u-\frac{u}{v_{gw}}\right)\right],
\end{split}
\end{equation}

\begin{equation}
\begin{split}
\eta(u)&=\{[\cos(u)-\cos(u\mu_1)][\cos(u)-\cos(u\mu_2)]\mu_1\mu_2 \\
&+[\sin(u)-\mu_1\sin(u\mu_1)][\sin(u)-\mu_2\sin(u\mu_2)]\}/
[(1-\mu_1^2)(1-\mu_2^2)],
\end{split}
\end{equation}
$\mu_1=\cos\theta_1/v_{gw}$, $\mu_2=\cos\theta_{2}/v_{gw}$, $\cos\theta_2=\cos\sigma\cos\theta_1+\sin\sigma\sin\theta_1\cos\epsilon$.
In the massless or high frequency limit, $v_{gw}=c$ and we recover those results in \cite{Liang:2019pry}. In the low frequency and massless limits,
$m_g\ll \omega\ll 2\pi f_*$, we get $R^+=R^\times=R^x=R^y=\sin^2\sigma/5$
and $R^b=R^l=\sin^2\sigma/15$. On the other hand, in the limit $\omega\to m_g$,
we get $R^A=0$. We show the results of the averaged response function for equal arm interferometric GW detectors without optical cavities in the arms in Fig. \ref{rufige}.
Since the observations constrain the mass of gravitons to be $m_g\le 5.0\times 10^{-23}$ eV \cite{LIGOScientific:2019fpa} and space-based GW detectors operate in the mHZ frequency,
we have $m_g\ll \omega$ and the effect of the propagation speed is negligible as shown in Fig. \ref{rufige}.

\section*{Averaged Response functions}
\label{appb}

\subsection{Beacon combinations $P$, $Q$, $R$}
\label{beacon}

The averaged response functions of $P$ are
\begin{equation}
\label{rptensor}
\begin{split}					
u^2R^+_P=&u^2R^\times_P=[1-\cos u_1\cos( u_2- u_3)]\left[\frac43-\frac{2}{ u_1^2}+\frac{\sin(2 u_1)}{u_1^3}\right]+\sin^2 u_1\left[\frac83\right.\\
&-\frac{2}{ u_2^2}-\frac{2}{ u_3^2}+\left.\frac{\sin(2 u_2)}{ u_2^3}+\frac{\sin(2 u_3)}{ u_3^3}\right]+\frac1{8\pi}\int_{-1}^1d\mu_2\int_0^{2\pi}d\epsilon\left\{\eta_P^0\right.\\
&+\left[1-\frac{2(g_3/g_1)^2\sin^2\sigma\sin^2\epsilon}{1-\mu_1^2}\right]\eta_P^1+\left[\frac{4(g_3/g_1)\kappa_1\kappa_2\sin^2\sigma\sin^2\epsilon}{(1-\mu_1^2)(1-\mu_3^2)}\right.\\
&\left.+\left(1-\frac{2(g_3/g_1)^2\sin^2\sigma\sin^2\epsilon}{1-\mu_1^2}\right)\left(1-\frac{2\sin^2\sigma\sin^2\epsilon}{1-\mu_3^2}\right)\right]\eta_P^2\\
&+\left.\left(1-\frac{2\sin^2\sigma\sin^2\epsilon}{1-\mu_3^2}\right)\eta_P^3\right\},
\end{split}
\end{equation}

\begin{equation}
\label{rpvector}
\begin{split}
u^2R^x_P=& u^2R^y_P=2\left[1-\cos u_1\cos( u_2- u_3)\right]\left[2\gamma_E-\frac{14}{3}+\frac{4}{ u_1^2}-2\bigg({\rm Ci}(2 u_1)-\ln(2 u_1)\bigg)\right.\\
&+\left.\left(\frac1{ u_1}-\frac{2}{ u_1^3}\right)\sin(2 u_1)\right]+2\sin^2 u_1\left[4\gamma_E-\frac{28}{3}+\frac{4}{ u_2^2}+\frac{4}{ u_3^2}-2\bigg({\rm Ci}(2 u_2)\right.\\
&+\left.{\rm Ci}(2 u_3)-\ln(4 u_2 u_3)\bigg)+\left(\frac{1}{ u_2}-\frac{2}{ u_2^3}\right)\sin(2 u_2)+\left(\frac{1}{ u_3}-\frac{2}{ u_3^3}\right)\sin(2 u_3)\right]\\
&+\frac1{2\pi}\int_{-1}^1d\mu_2\int_0^{2\pi}d\epsilon\left\{\frac{\mu_1^2}{1-\mu_1^2}\eta_\zeta^0+\frac{\mu_1\mu_2\kappa_2}{(1-\mu_1^2)\sqrt{1-\mu_2^2}}\eta_\zeta^1\right.\\
&+\left.\frac{\left[\kappa_1\kappa_2+(g_3/g_1)\sin^2\sigma\sin^2\epsilon\right]\mu_1\mu_3}
{(1-\mu_1^2)(1-\mu_3^2)}\eta_\zeta^2+\frac{\mu_2\mu_3\kappa_1}{(1-\mu_3^2)\sqrt{1-\mu_2^2}}\eta_\zeta^3\right\},
\end{split}
\end{equation}

\begin{equation}
\label{rpbeq}
\begin{split}
u^2R^b_P=&2[1-\cos u_1\cos( u_2- u_3)]\left[\frac43-\frac{2}{ u_1^2}+\frac{\sin(2 u_1)}{ u_1^3}\right]+2\sin^2 u_1\left[\frac83-\frac{2}{ u_2^2}-\frac{2}{ u_3^2}\right.\\
&+\left.\frac{\sin(2 u_2)}{ u_2^3}+\frac{\sin(2 u_3)}{ u_3^3}\right]
+\frac1{4\pi}\int_{-1}^1d\mu_2\int_0^{2\pi}d\epsilon\left\{\eta_P^0+\eta_P^1+\eta_P^2+\eta_P^3\right\},
\end{split}
\end{equation}

\begin{equation}
\label{rpleq}
\begin{split}
u^2R^l_P=&[1-\cos u_1\cos( u_2- u_3)]\left[\frac{37}{6}-4\gamma_E-\frac{4}{ u_1^2}+\frac12\cos(2 u_1)+\left(\frac{2}{ u_1^3}-\frac{2}{ u_1}\right)\sin(2 u_1)\right.\\
&+\left.4\bigg({\rm Ci}(2 u_1)-\ln(2 u_1)\bigg)+ u_1{\rm Si}(2 u_1)\right]+\sin^2 u_1\left[\frac{37}{3}-8\gamma_E-\frac{4}{ u_2^2}-\frac{4}{ u_3^2}\right.\\
&+\frac12\bigg(\cos(2 u_2)+\cos(2 u_3)\bigg)+\left(\frac{2}{ u_2^3}-\frac{2}{ u_2}\right)\sin(2 u_2)+\left(\frac{2}{ u_3^3}-\frac{2}{ u_3}\right)\sin(2 u_3)\\
&+\left.4\bigg({\rm Ci}(2 u_2)+{\rm Ci}(2 u_3)-\ln(4 u_2 u_3)\bigg)+ u_2{\rm Si}(2 u_2)+ u_3{\rm Si}(2 u_3)\right]\\
&+\frac1{4\pi}\int_{-1}^1d\mu_2\int_0^{2\pi}d\epsilon\left\{\frac{\mu_1^4}{(1-\mu_1^2)^2}\eta_P^0+\frac{\mu_1^2\mu_2^2}{(1-\mu_1^2)(1-\mu_2^2)}\eta_P^1\right.\\
&+\left.\frac{\mu_1^2\mu_3^2}{(1-\mu_1^2)(1-\mu_3^2)}\eta_P^2+\frac{\mu_2^2\mu_3^2}{(1-\mu_2^2)(1-\mu_3^2)}\eta_P^3\right\},
\end{split}	
\end{equation}
where
\begin{gather}
\begin{split}
\eta^0_P=2(1-\mu_1^2)[\cos u_1-\cos( u_2- u_3)][\cos( u_1\mu_1)-\cos u_1]\cos( u_1\mu_1+ u_2\mu_2- u_3\mu_3)
\end{split}\\
\begin{split}
\eta_P^1=&8(1+\mu_2)\sin u_1\sin\frac12( u_2- u_2\mu_2)\left[(1+\mu_1)\sin\frac12( u_1- u_1\mu_1)\sin\frac12( u_1- u_2+ u_3)\times\right.\\
&\cos\frac12( u_1\mu_1+ u_2\mu_2+ u_3-2 u_3\mu_3)-\\
&\left.(1-\mu_1)\sin\frac12( u_1+ u_1\mu_1)\sin\frac12( u_1+ u_2- u_3)\cos\frac12( u_1\mu_1+ u_2\mu_2- u_3)\right]
\end{split}\\
\begin{split}
\eta^2_P=&8(1+\mu_3)\sin u_1\sin\frac12( u_3- u_3\mu_3)\left[(1-\mu_1)\sin\frac12( u_1+ u_1\mu_1)\sin\frac12( u_1+ u_2- u_3)\times\right.\\
&\cos\frac12( u_1\mu_1- u_2+2 u_2\mu_2- u_3\mu_3)-\\
&\left.(1+\mu_1)\sin\frac12( u_1- u_1\mu_1)\sin\frac12( u_1- u_2+ u_3)\cos\frac12( u_1\mu_1+ u_2- u_3\mu_3)\right]
\end{split}\\
\begin{split}
\eta^3_P=&-8(1+\mu_2)(1+\mu_3)\sin^2 u_1\sin\frac12( u_2- u_2\mu_2)\sin\frac12( u_3- u_3\mu_3)\cos\frac12( u_2- u_2\mu_2\\
&- u_3+ u_3\mu_3)
\end{split}
\end{gather}

Similarly, we can obtain the averaged response functions $R^A_Q$ and $R^A_R$ through the cyclic permutation $ u_1\to u_2\to u_3\to u_1$.

\subsection{Monitor combinations}
\label{monitor}

The averaged response functions of $E$ are
\begin{equation}
\label{retensor}
\begin{split}
u^2R^+_E=& u^2R^\times_E=[1-\cos u_1\cos( u_2- u_3)]\left[\frac43-\frac2{ u_1^2}+\frac{\sin(2 u_1)}{ u_1^3}\right]+\sin^2 u_1\left [\frac83\right.\\
&\left.-\frac{2}{ u_2^2}-\frac{2}{ u_3^2}+\frac{\sin(2 u_2)}{ u_2^3}+\frac{\sin(2 u_3)}{ u_3^3}\right]+\frac1{8\pi}\int_{-1}^1d\mu_2\int_0^{2\pi}d\epsilon\left\{\eta_E^0\right.\\
&+\left[1-\frac{2(g_3/g_1)^2\sin^2\sigma\sin^2\epsilon}{1-\mu_1^2}\right]\eta^1_E+\left[\frac{4(g_3/g_1)\kappa_1\kappa_2\sin^2\sigma\sin^2\epsilon}{(1-\mu_1^2)(1-\mu_3^2)}\right.\\
&\left.+\left(1-\frac{2(g_3/g_1)^2\sin^2\sigma\sin^2\epsilon}{1-\mu_1^2}\right)\left(1-\frac{2\sin^2\sigma\sin^2\epsilon}{1-\mu_3^2}\right)\right]\eta_E^2\\
&\left.+\left(1-\frac{2\sin^2\sigma\sin^2\epsilon}{1-\mu_3^2}\right)\eta_E^3\right\},
\end{split}
\end{equation}

\begin{equation}
\label{revector}
\begin{split}
u^2R^x_E=& u^2R^y_E=2[1-\cos u_1\cos( u_2- u_3)]\left[2\gamma_E-\frac{14}{3}+\frac{4}{ u_1^2}-2\bigg({\rm Ci}(2 u_1)-\ln(2 u_1)\bigg)\right.\\
&+\left.\left(\frac{1}{ u_1}-\frac{2}{ u_1^3}\right)\sin(2 u_1)\right]+2\sin^2 u_1\left[4\gamma_E-\frac{28}{3}+\frac{4}{ u_2^2}+\frac{4}{ u_3^2}-2\bigg({\rm Ci}(2 u_2)\right.\\
&\left.+{\rm Ci}(2 u_3)-\ln(4 u_2 u_3)\bigg)+\left(\frac{1}{ u_2}-\frac{2}{ u_2^2}\right)\sin(2 u_2)+\left(\frac{1}{ u_3}-\frac{2}{ u_3^2}\right)\sin(2 u_3)\right]\\
&+\frac1{2\pi}\int_{-1}^1d\mu_2\int_0^{2\pi}d\epsilon\left\{\frac{\mu_1^2}{1-\mu_1^2}\eta_E^0+\frac{\mu_1\mu_2\kappa_2}{(1-\mu_1^2)\sqrt{1-\mu_2^2}}\eta_E^1\right.\\
&+\left.\frac{\left[\kappa_1\kappa_2+(g_3/g_1)\sin^2\sigma\sin^2\epsilon\right]\mu_1\mu_3}{(1-\mu_1^2)
	(1-\mu_3^2)}\eta_E^2+\frac{\mu_2\mu_3\kappa_1}{(1-\mu_3^2)\sqrt{1-\mu_2^2}}\eta_E^3\right\},
\end{split}
\end{equation}

\begin{equation}
\label{rebeq}
\begin{split}
u^2R^b_E=&2[1-\cos u_1\cos( u_2- u_3)]\left[\frac43-\frac{2}{ u_1^2}+\frac{\sin(2 u_1)}{ u_1^3}\right]+2\sin^2 u_1\left[\frac83-\frac{2}{ u_2^2}-\frac{2}{ u_3^2}\right.\\
&+\left.\frac{\sin(2 u_2)}{ u_2^3}+\frac{\sin(2 u_3)}{ u_3^3}\right]
+\frac1{4\pi}\int_{-1}^1d\mu_2\int_0^{2\pi}d\epsilon\{\eta_E^0+\eta_E^1+\eta_E^2+\eta_E^3\},
\end{split}
\end{equation}

\begin{equation}
\label{releq}
\begin{split}
u^2R^l_E=&[1-\cos u_1\cos( u_2- u_3)]\left[\frac{37}{6}-4\gamma_E-\frac{4}{ u_1^2}+\frac12\cos(2 u_1)+\left(\frac{2}{ u_1^3}-\frac{2}{ u_1}\right)\sin2 u_1\right.\\
&+\left.4\bigg({\rm Ci}(2 u_1)-\ln(2 u_1)\bigg)+ u_1{\rm Si}(2 u_1)\right]+\sin^2 u_1\left[\frac{37}{3}-8\gamma_E-\frac{4}{ u_2^2}-\frac{4}{ u_3^2}\right.\\
&+\frac{1}{2}\bigg(\cos(2 u_2)+\cos(2 u_3)\bigg)+4\bigg({\rm Ci}(2 u_2)+{\rm Ci}(2 u_3)-\ln(4 u_2 u_3)\bigg)\\
&-\left.\left(\frac{2}{ u_2}-\frac{2}{ u_2^3}\right)\sin(2 u_2)-\left(\frac{2}{ u_3}-\frac{2}{ u_3^3}\right)\sin(2 u_3)+ u_2{\rm Si}(2 u_2)+ u_3{\rm Si}(2 u_3)\right]\\
&+\frac1{4\pi}\int_{-1}^1d\mu_2\int_0^{2\pi}d\epsilon\left\{\frac{\mu_1^4}{(1-\mu_1^2)^2}\eta_E^0+\frac{\mu_1^2\mu_2^2}{(1-\mu_1^2)(1-\mu_2^2)}\eta_E^1\right.\\
&+\left.\frac{\mu_1^2\mu_3^2}{(1-\mu_1^2)(1-\mu_3^2)}\eta_E^2+\frac{\mu_2^2\mu_3^2}{(1-\mu_2^2)(1-\mu_3^2)}\eta_E^3\right\},
\end{split}
\end{equation}
where
\begin{equation}
\eta^0_E=2(1-\mu_1^2)[\cos u_1-\cos( u_2- u_3)][\cos( u_1\mu_1)-\cos u_1]\cos( u_1\mu_1+ u_2\mu_2- u_3\mu_3),
\end{equation}

\begin{equation}
\begin{split}
\eta_E^1=&8(1-\mu_2)\sin u_1\sin\frac12( u_2+ u_2\mu_2)\left[(1-\mu_1)\sin\frac12( u_1+ u_1\mu_1)\sin\frac12( u_1- u_2\right.\\
&+ u_3)\cos\frac12( u_1\mu_1+ u_2\mu_2+ u_3)-(1+\mu_1)\sin\frac12( u_1- u_1\mu_1)\sin\frac12( u_1+ u_2\\
&\left.- u_3)\cos\frac12( u_1\mu_1+ u_2\mu_2- u_3-2 u_3\mu_3)\right]
\end{split}
\end{equation}

\begin{equation}
\begin{split}
\eta^2_E=&8(1-\mu_3)\sin u_1\sin\frac12( u_3+ u_3\mu_3)\left[(1+\mu_1)\sin\frac12(u_1- u_1\mu_1)\sin\frac12( u_1+ u_2\right.\\
&-u_3)\cos\frac12( u_1\mu_1- u_2- u_3\mu_3)-(1-\mu_1)\sin\frac12( u_1+ u_1\mu_1)\sin\frac12( u_1- u_2\\
&\left.+ u_3)\cos\frac12( u_1\mu_1+ u_2+2 u_2\mu_2- u_3\mu_3)\right]
\end{split}
\end{equation}

\begin{equation}
\begin{split}
\eta^3_E=&-8(1-\mu_2)(1-\mu_3)\sin^2 u_1\sin\frac12( u_2+ u_2\mu_2)\sin\frac12( u_3+ u_3\mu_3)\cos\frac12( u_2+ u_2\mu_2\\
&- u_3- u_3\mu_3)
\end{split}
\end{equation}

Similarly, we can obtain the averaged response functions $R^A_F$ and $R^A_G$ through the cyclic permutation $ u_1\to u_2\to u_3\to u_1$.

\subsection{Relay combinations}
\label{relay}

The averaged response functions of $U$ are
\begin{equation}
\label{rutensor}
\begin{split}
u^2R^+_U=& u^2R^\times_U=[1-\cos u_1\cos( u_2+ u_3)]\left[\frac43-\frac2{ u_1^2}+\frac{\sin(2 u_1)}{ u_1^3}\right]+\sin^2 u_1\left[\frac83\right.\\
&\left.-\frac{2}{ u_2^2}-\frac{2}{ u_3^2}+\frac{\sin(2 u_2)}{ u_2^3}+\frac{\sin(2 u_3)}{ u_3^3}\right]+\frac1{8\pi}\int_{-1}^1d\mu_2\int_0^{2\pi}d\epsilon\left\{\eta_U^0\right.\\
&+\left[1-\frac{2(g_3/g_1)^2\sin^2\sigma\sin^2\epsilon}{1-\mu_1^2}\right]\eta_U^1+\left[\frac{4(g_3/g_1)\kappa_1\kappa_2\sin^2\sigma\sin^2\epsilon}{(1-\mu_1^2)(1-\mu_3^2)}\right.\\
&\left.+\left(1-\frac{2(g_3/g_1)^2\sin^2\sigma\sin^2\epsilon}{1-\mu_1^2}\right)\left(1-\frac{2\sin^2\sigma\sin^2\epsilon}{1-\mu_3^2}\right)\right]\eta_U^2\\
&\left.+\left(1-\frac{2\sin^2\sigma\sin^2\epsilon}{1-\mu_3^2}\right)\eta_U^3\right\},
\end{split}
\end{equation}

\begin{equation}
\label{ruvector}
\begin{split}
u^2R^x_U=& u^2R^y_U=2[1-\cos u_1\cos( u_2+ u_3)]\left[2\gamma_E-\frac{14}{3}+\frac{4}{ u_1^2}-2\bigg({\rm Ci}(2 u_1)-\ln(2 u_1)\bigg)\right.\\
&+\left.\left(\frac{1}{ u_1}-\frac{2}{ u_1^3}\right)\sin(2 u_1)\right]+2\sin^2 u_1\left[4\gamma_E-\frac{28}{3}+\frac{4}{ u_2^2}+\frac{4}{ u_3^2}-2\bigg({\rm Ci}(2 u_2)\right.\\
&\left.+{\rm Ci}(2 u_3)-\ln(4 u_2 u_3)\bigg)+\left(\frac{1}{ u_2}-\frac{2}{ u_2^3}\right)\sin(2 u_2)+\left(\frac{1}{ u_3}-\frac{2}{ u_3^3}\right)\sin(2 u_3)\right]\\
&+\frac1{2\pi}\int_{-1}^1d\mu_2\int_0^{2\pi}d\epsilon\left\{\frac{\mu_1^2}{1-\mu_1^2}\eta_U^0+\frac{\mu_1\mu_2\kappa_2}{(1-\mu_1^2)\sqrt{1-\mu_2^2}}\eta_U^1\right.\\
&+\left.\frac{\left[\kappa_1\kappa_2+(g_3/g_1)\sin^2\sigma\sin^2\epsilon\right]\mu_1\mu_3}{(1-\mu_1^2)(1-\mu_3^2)}\eta_U^2
+\frac{\mu_2\mu_3\kappa_1}{(1-\mu_3^2)\sqrt{1-\mu_2^2}}\eta_U^3\right\},
\end{split}
\end{equation}

\begin{equation}
\label{rubeq}
\begin{split}
u^2R^b_U=&2[1-\cos u_1\cos( u_2+ u_3)]\left[\frac43-\frac{2}{ u_1^2}+\frac{\sin(2 u_1)}{ u_1^3}\right]+2\sin^2 u_1\left[\frac83-\frac{2}{ u_2^2}-\frac{2}{ u_3^2}\right.\\
&+\left.\frac{\sin(2 u_2)}{ u_2^3}+\frac{\sin(2 u_3)}{ u_3^3}\right]
+\frac1{4\pi}\int_{-1}^1d\mu_2\int_0^{2\pi}d\epsilon\left\{\eta_U^0+\eta_U^1+\eta_U^2+\eta_U^3\right\},
\end{split}
\end{equation}

\begin{equation}
\label{ruleq}
\begin{split}
u^2R^l_U=&[1-\cos u_1\cos( u_2+ u_3)]\left[\frac{37}{6}-4\gamma_E-\frac{4}{ u_1^2}+\frac12\cos2 u_1-(\frac{2}{ u_1}-\frac{2}{ u_1^3})\sin2 u_1\right.\\
&\left.+4\bigg({\rm Ci}(2 u_1)-\ln(2 u_1)\bigg)+ u_1{\rm Si}(2 u_1)\right]+\sin^2 u_1\left[\frac{37}{3}-8\gamma_E-\frac{4}{ u_2^2}-\frac{4}{ u_3^2}\right.\\
&+\left.\frac{1}{2}\bigg(\cos(2 u_2)+\cos(2 u_3)\bigg)+4\bigg({\rm Ci}(2 u_2)+{\rm Ci}(2 u_3)-\ln(4 u_2 u_3)\bigg)\right.\\
&\left.-(\frac{2}{ u_2}-\frac{2}{ u_2^3})\sin2 u_2-\left(\frac{2}{ u_3}-\frac{2}{ u_3^3}\right)\sin(2 u_3)+ u_2{\rm Si}(2 u_2)+ u_3{\rm Si}(2 u_3)\right]\\
&+\frac1{4\pi}\int_{-1}^1d\mu_2\int_0^{2\pi}d\epsilon\left\{\frac{\mu_1^4}{(1-\mu_1^2)^2}\eta_U^0+\frac{\mu_1^2\mu_2^2}{(1-\mu_1^2)(1-\mu_2^2)}\eta_U^1\right.\\
&\left.+\frac{\mu_1^2\mu_3^2}{(1-\mu_1^2)(1-\mu_3^2)}\eta_U^2+\frac{\mu_2^2\mu_3^2}{(1-\mu_2^2)(1-\mu_3^2)}\eta_U^3\right\}
\end{split}
\end{equation}
where
\begin{gather}
\begin{split}
\eta^0_U=2(1-\mu_1^2)[\cos u_1-\cos( u_2+ u_3)][\cos( u_1\mu_1)-\cos u_1]\cos( u_1\mu_1+ u_2\mu_2- u_3\mu_3)\nonumber
\end{split}\\
\begin{split}
\eta_U^1=&8(1-\mu_2)\sin u_1\sin\frac12( u_2+ u_2\mu_2)\left[(1-\mu_1)\sin\frac12( u_1+ u_1\mu_1)\sin\frac12( u_1- u_2- u_3)\times\right.\\
&\cos\frac12( u_1\mu_1+ u_2\mu_2- u_3)-\\
&\left.(1+\mu_1)\sin\frac12( u_1- u_1\mu_1)\sin\frac12( u_1+ u_2+ u_3)\cos\frac12( u_1\mu_1+ u_2\mu_2+ u_3-2 u_3\mu_3)\right]
\end{split}\\
\begin{split}
\eta^2_U=&8(1+\mu_3)\sin u_1\sin\frac12( u_3- u_3\mu_3)\left[(1-\mu_1)\sin\frac12( u_1+ u_1\mu_1)\sin\frac12( u_1- u_2- u_3)\times\right.\\
&\cos\frac12( u_1\mu_1+ u_2+2 u_2\mu_2- u_3\mu_3)-\\
&\left.(1+\mu_1)\sin\frac12( u_1- u_1\mu_1)\sin\frac12( u_1+ u_2+ u_3)\cos\frac12( u_1\mu_1- u_2- u_3\mu_3)\right]
\end{split}\\
\begin{split}
\eta^3_U=&8(1-\mu_2)(1+\mu_3)\sin^2 u_1\sin\frac12( u_2+ u_2\mu_2)\sin\frac12( u_3- u_3\mu_3)\cos\frac12( u_2+ u_2\mu_2\\
&+ u_3- u_3\mu_3)
\end{split}
\end{gather}

Similarly, we can obtain the averaged response functions $R^A_V$ and $R^A_W$ through the cyclic permutation $ u_1\to u_2\to u_3\to u_1$.

\subsection{The $\alpha$, $\beta$, $\gamma$ (six-pulse) combinations}
\label{alpha}

The averaged response functions of $\alpha$ are
\begin{equation}
\label{ratensor}
\begin{split}
u^2R^+_\alpha=& u^2R_\alpha^\times=2-\frac1{ u_1^2}-\frac1{ u_2^2}-\frac1{ u_3^2}+\frac{\sin(2 u_1)}{2 u_1^3}+\frac{\sin(2 u_2)}{2 u_2^3}+\frac{\sin(2 u_3)}{2 u_3^3}+\\
&\left[\left(\frac1{ u_2^2}+\frac13\right)\cos u_2-\frac{\sin u_2}{ u_2^3}\right]\cos( u_1+ u_3)+\left[\left(\frac1{ u_3^2}+\frac13\right)\cos u_3-\frac{\sin u_3}{ u_3^3}\right]\times\\
&\cos( u_1+u_2)+\frac1{8\pi}\int_{-1}^1d\mu_2\int_0^{2\pi}d\epsilon\left\{\eta_\alpha^0+\left[1-\frac{2(g_3/g_1)^2\sin^2\sigma\sin^2\epsilon}{1-\mu_1^2}\right]\eta_\alpha^1\right.\\
&+\left[\left(1-\frac{2(g_3/g_1)^2\sin^2\sigma\sin^2\epsilon}{1-\mu_1^2}\right)\left(1-\frac{2\sin^2\sigma\sin^2\epsilon}{1-\mu_3^2}\right)\right.\\
&\left.+\frac{4(g_3/g_1)\kappa_1\kappa_2\sin^2\sigma\sin^2\epsilon}{(1-\mu_1^2)(1-\mu_3^2)}\right]\eta_\alpha^2+\left.\left(1-\frac{2\sin^2\sigma\sin^2\epsilon}{1-\mu_3^2}\right)\eta_\alpha^3\right\},
\end{split}
\end{equation}

\begin{equation}
\label{ravector}
\begin{split}
u^2R_\alpha^x=& u^2R_\alpha^y=6\gamma_E-14+4\left(\frac1{ u_1^2}+\frac1{ u_2^2}+\frac1{ u_3^2}\right)-2\left[{\rm Ci}(2 u_1)+{\rm Ci}(2 u_2)+{\rm Ci}(2 u_3)\right.\\
&\left.-\ln(8 u_1 u_2 u_3)\right]+\left(\frac1{ u_1}-\frac2{ u_1^3}\right)\sin2 u_1+\left(\frac1{ u_2}-\frac2{ u_2^3}\right)\sin2 u_2+\\
&\left(\frac1{ u_3}-\frac2{ u_3^3}\right)\sin2 u_3+2\cos( u_1+ u_3)\left[\left(\frac13-\frac2{ u_2^2}\right)\cos u_2-\left(\frac1{ u_2}-\frac2{ u_2^3}\right)\sin u_2\right]\\
&+2\cos( u_1+ u_2)\left[\left(\frac13-\frac2{ u_3^2}\right)\cos u_3-\left(\frac1{ u_3}-\frac2{ u_3^3}\right)\sin u_3\right]\\
&+\frac1{2\pi}\int_{-1}^1d\mu_2\int_0^{2\pi}d\epsilon\left\{\frac{\mu_1^2}{1-\mu_1^2}\eta_\alpha^0+\frac{\mu_1\mu_2\kappa_2}{(1-\mu_1^2)\sqrt{1-\mu_2^2}}\eta_\alpha^1\right.\\
&\left.+\frac{[\kappa_1\kappa_2+(g_3/g_1)\sin^2\sigma\sin^2\epsilon]\mu_1\mu_3}{(1-\mu_1^2)(1-\mu_3^2)}\eta_\alpha^2
+\frac{\mu_2\mu_3\kappa_1}{(1-\mu_3^2)\sqrt{1-\mu_2^2}}\eta_\alpha^3\right\},
\end{split}
\end{equation}

\begin{equation}
\label{rabeq}
\begin{split}
u^2R_\alpha^b=&4-2\left(\frac1{ u_1^2}+\frac1{ u_2^2}+\frac1{ u_3^2}\right)+\frac{\sin(2 u_1)}{ u_1^3}+\frac{\sin(2 u_2)}{ u_2^3}+\frac{\sin(2 u_3)}{ u_3^3}\\
&+2\left[\left(\frac{1}{ u_2^2}+\frac13\right)\cos u_2-\frac1{ u_2^3}\sin u_2\right]\cos( u_1+ u_3)\\
&+2\left[\left(\frac{1}{ u_3^2}+\frac13\right)\cos u_3-\frac1{ u_3^3}\sin u_3\right]\cos( u_1+ u_2)\\
&+\frac1{4\pi}\int_{-1}^1d\mu_2\int_0^{2\pi}d\epsilon\left\{\eta_\alpha^0+\eta_\alpha^1+\eta_\alpha^2+\eta_\alpha^3\right\},
\end{split}
\end{equation}

\begin{equation}
\label{raleq}
\begin{split}
u^2R_\alpha^l&=\frac{37}{4}-6\gamma_E-2\left(\frac1{ u_1^2}+\frac1{ u_2^2}+\frac1{ u_3^2}\right)+\frac14\left[\cos(2 u_1)+\cos(2 u_2)+\cos(2 u_3)\right]\\
&+\left(\frac1{ u_1^3}-\frac1{ u_1}\right)\sin(2 u_1)+\left(\frac1{ u_2^3}-\frac1{ u_2}\right)\sin(2 u_2)+\left(\frac1{ u_3^3}-\frac1{ u_3}\right)\sin(2 u_3)\\
&+2\left[{\rm Ci}(2 u_1)+{\rm Ci}(2 u_2)+{\rm Ci}(2 u_3)-\ln(8 u_1 u_2 u_3)\right]+\frac12\left[ u_1{\rm Si}(2 u_1)+ u_2{\rm Si}(2 u_2)\right.\\
&\left.+ u_3{\rm Si}(2 u_3)\right]+\left[\frac12\gamma_E-\frac43+\frac2{ u_2^2}-\frac12\bigg({\rm Ci}(2 u_2)-\ln(2 u_2)\bigg)\right]\cos( u_1+ u_3)\cos u_2\\
&+\left[\frac2{ u_2}-\frac2{ u_2^3}-\frac12{\rm Si}(2 u_2)\right]\cos( u_1+ u_3)\sin u_2\\
&+\left[\frac12\gamma_E-\frac43+\frac2{ u_3^2}-\frac12\bigg({\rm Ci}(2 u_3)-\ln(2 u_3)\bigg)\right]\cos( u_1+ u_2)\cos u_3\\
&+\left[\frac2{ u_3}-\frac2{ u_3^3}-\frac12{\rm Si}(2 u_3)\right]\cos( u_1+ u_2)\sin u_3\\
&+\frac1{4\pi}\int_{-1}^1d\mu_2\int_0^{2\pi}d\epsilon\left\{\frac{\mu_1^4}{(1-\mu_1^2)^2}\eta_\alpha^0+\frac{\mu_1^2\mu_2^2}{(1-\mu_1^2)(1-\mu_2^2)}\eta_\alpha^1\right.\\
&\left.+\frac{\mu_1^2\mu_3^2}{(1-\mu_1^2)(1-\mu_3^2)}\eta_\alpha^2+\frac{\mu_2^2\mu_3^2}{(1-\mu_2^2)(1-\mu_3^2)}\eta_\alpha^3\right\},
\end{split}
\end{equation}
where

\begin{gather}
\begin{split}
\eta_\alpha^0=&\left(1-\mu_1^2\right)\left[\cos u_1-\cos( u_1\mu_1)\right]\cos\left( u_1\mu_1+ u_2+ u_2\mu_2- u_3- u_3\mu_3\right)
\end{split}\\
\begin{split}
\eta_\alpha^1=&2(1+\mu_1)\sin\frac12( u_1- u_1\mu_1)\left[(1+\mu_2)\sin\frac12( u_2- u_2\mu_2)\cos\frac12( u_1+ u_1\mu_1+ u_2\right.\\
&+\left.u_2\mu_2-2 u_3\mu_3)-(1-\mu_2)\sin\frac12( u_2+ u_2\mu_2)\cos\frac12( u_1- u_1\mu_1- u_2- u_2\mu_2\right.\\
&\left.+2 u_3+2 u_3\mu_3)\right]-2(1-\mu_1)\sin\frac12( u_1+ u_1\mu_1)\left[(1+\mu_2)\sin\frac12( u_2- u_2\mu_2)\right.\\
&\left.\times\cos\frac12( u_1- u_1\mu_1- u_2- u_2\mu_2+2 u_3)-(1-\mu_2)\sin\frac12( u_2+ u_2\mu_2)\cos\frac12( u_1\right.\\
&\left.+ u_1\mu_1+ u_2+ u_2\mu_2)\right]
\end{split}\\
\begin{split}
\eta_\alpha^2=&2(1-\mu_1)\sin\frac12( u_1+ u_1\mu_1)\left[(1+\mu_3)\sin\frac12( u_3- u_3\mu_3)\cos\frac12( u_1- u_1\mu_1-2 u_2\mu_2\right.\\
&+ u_3+ u_3\mu_3)-(1-\mu_3)\sin\frac12( u_3+ u_3\mu_3)\cos\frac12( u_1+ u_1\mu_1+2 u_2+2 u_2\mu_2- u_3-\\
&\left. u_3\mu_3)\right]-2(1+\mu_1)\sin\frac12( u_1- u_1\mu_1)\left[(1+\mu_3)\sin\frac12( u_3- u_3\mu_3)\cos\frac12( u_1+ u_1\mu_1\right.\\
&\left.+2 u_2- u_3- u_3\mu_3)-(1-\mu_3)\sin\frac12( u_3+ u_3\mu_3)\cos\frac12( u_1- u_1\mu_1+ u_3+ u_3\mu_3)\right]
\end{split}\\
\begin{split}
\eta_\alpha^3=&2(1-\mu_2)\sin\frac12( u_2+ u_2\mu_2)\left[(1+\mu_3)\sin\frac12( u_3- u_3\mu_3)\cos\frac12(2 u_1+ u_2- u_2\mu_2\right.\\
&\left.+ u_3+ u_3\mu_3)-(1-\mu_3)\sin\frac12( u_3+ u_3\mu_3)\cos\frac12( u_2+ u_2\mu_2- u_3- u_3\mu_3)\right]-\\
&2(1+\mu_2)\sin\frac12( u_2- u_2\mu_2)\left[(1+\mu_3)\sin\frac12( u_3- u_3\mu_3)\cos\frac12( u_2- u_2\mu_2- u_3\right.\\
&\left.+ u_3\mu_3)-(1-\mu_3)\sin\frac12( u_3+ u_3\mu_3)\cos\frac12(2 u_1+ u_2+ u_2\mu_2+ u_3- u_3\mu_3)\right]
\end{split}
\end{gather}

Similarly, we can obtain the averaged response functions $R^A_\beta$ and $R^A_\gamma$ through the cyclic permutation $ u_1\to u_2\to u_3\to u_1$.

\subsection{Fully symmetric (Sagnac) combination}
\label{sagnac}

The averaged response functions of $\zeta$ are
\begin{equation}
\label{rztensor}
\begin{split}
u^2R_\zeta^+&= u^2R_\zeta^\times=2-\frac1{ u_1^2}-\frac1{ u_2^2}-\frac1{ u_3^2}+\frac{\sin(2 u_1)}{2 u_1^3}+\frac{\sin(2 u_2)}{2 u_2^3}+\frac{\sin(2 u_3)}{2 u_3^3}+\\
&\left[\left(\frac1{ u_2^2}+\frac13\right)\cos u_2-\frac{\sin u_2}{ u_2^3}\right]\cos( u_1- u_3)+\left[\left(\frac1{ u_3^2}+\frac13\right)\cos u_3-\frac{\sin u_3}{ u_3^3}\right]\\
&\times\cos( u_1- u_2)+\frac1{8\pi}\int_{-1}^1d\mu_2\int_0^{2\pi}d\epsilon\left\{\eta_\zeta^0+\left[1-\frac{2(g_3/g_1)^2\sin^2\sigma\sin^2\epsilon}{1-\mu_1^2}\right]\eta_\zeta^1\right.\\
&+\left[\left(1-\frac{2(g_3/g_1)^2\sin^2\sigma\sin^2\epsilon}{1-\mu_1^2}\right)\left(1-\frac{2\sin^2\sigma\sin^2\epsilon}{1-\mu_3^2}\right)\right.\\
&\left.\left.+\frac{4(g_3/g_1)\kappa_1\kappa_2\sin^2\sigma\sin^2\epsilon}{(1-\mu_1^2)(1-\mu_3^2)}\right]\eta_\zeta^2+\left(1-\frac{2\sin^2\sigma\sin^2\epsilon}{1-\mu_3^2}\right)\eta_\zeta^3\right\},
\end{split}
\end{equation}

\begin{equation}
\label{rzvector}
\begin{split}
u^2R_\zeta^x&= u^2R_\zeta^y=6\gamma_E-14+4\left(\frac1{ u_1^2}+\frac1{ u_2^2}+\frac1{ u_3^2}\right)-2\left[{\rm Ci}(2 u_1)+{\rm Ci}(2 u_2)+{\rm Ci}(2 u_3)\right.\\
&-\left.\ln(8 u_1 u_2 u_3)\right]+\left(\frac1{ u_1}-\frac2{ u_1^3}\right)\sin(2 u_1)+\left(\frac1{ u_2}-\frac2{ u_2^3}\right)\sin(2 u_2)+\\
&\left(\frac1{ u_3}-\frac2{ u_3^3}\right)\sin(2 u_3)+2\left[\left(\frac13-\frac2{ u_2^2}\right)\cos u_2-\left(\frac1{ u_2}-\frac2{ u_2^3}\right)\sin u_2\right]\cos( u_1- u_3)\\
&+2\left[\left(\frac13-\frac2{ u_3^2}\right)\cos u_3-\left(\frac1{ u_3}-\frac2{ u_3^3}\right)\sin u_3\right]\cos( u_1- u_2)\\
&+\frac1{2\pi}\int_{-1}^1d\mu_2\int_0^{2\pi}d\epsilon\left\{\frac{\mu_1^2}{1-\mu_1^2}\eta_\zeta^0+\frac{\mu_1\mu_2\kappa_2}{(1-\mu_1^2)\sqrt{1-\mu_2^2}}\eta_\zeta^1\right.\\
&\left.
+\frac{\left[\kappa_1\kappa_2+(g_3/g_1)\sin^2\sigma\sin^2\epsilon\right]\mu_1\mu_3}{(1-\mu_1^2)(1-\mu_3^2)}\eta_\zeta^2
+\frac{\mu_2\mu_3\kappa_1}{(1-\mu_3^2)\sqrt{1-\mu_2^2}}\eta_\zeta^3\right\},
\end{split}
\end{equation}

\begin{equation}
\label{rzbeq}
\begin{split}
u^2R_\zeta^b&=4-2\left(\frac1{ u_1^2}+\frac1{ u_2^2}+\frac1{ u_3^2}\right)+\frac{\sin(2 u_1)}{ u_1^3}+\frac{\sin(2 u_2)}{ u_2^3}+\frac{\sin(2 u_3)}{ u_3^3}\\
&+2\left[\left(\frac{1}{ u_2^2}+\frac13\right)\cos u_2-\frac1{ u_2^3}\sin u_2\right]\cos( u_1- u_3)\\
&+2\left[\left(\frac{1}{ u_3^2}+\frac13\right)\cos u_3-\frac1{ u_3^3}\sin u_3\right]\cos( u_1- u_2)\\
&+\frac1{4\pi}\int_{-1}^1d\mu_2\int_0^{2\pi}d\epsilon\left\{\eta_\zeta^0+\eta_\zeta^1+\eta_\zeta^2+\eta_\zeta^3\right\},
\end{split}
\end{equation}

\begin{equation}
\label{rzleq}
\begin{split}
u^2R_\zeta^l&=\frac{37}{4}-6\gamma_E-2\left(\frac1{ u_1^2}+\frac1{ u_2^2}+\frac1{ u_3^2}\right)+\frac14\left[\cos(2 u_1)+\cos(2 u_2)+\cos(2 u_3)\right]\\
&+\left(\frac1{ u_1^3}-\frac1{ u_1}\right)\sin(2 u_1)+\left(\frac1{ u_2^3}-\frac1{ u_2}\right)\sin(2 u_2)+\left(\frac1{ u_3^3}-\frac1{ u_3}\right)\sin(2 u_3)\\
&+2\left[{\rm Ci}(2 u_1)+{\rm Ci}(2 u_2)+{\rm Ci}(2 u_3)-\ln(8 u_1 u_2 u_3)\right]+\frac12\left[ u_1{\rm Si}(2 u_1)+ u_2{\rm Si}(2 u_2)\right.\\
&\left.+ u_3{\rm Si}(2 u_3)\right]+\left[\frac12\gamma_E-\frac43+\frac2{ u_2^2}-\frac12\bigg({\rm Ci}(2 u_2)-\ln(2 u_2)\bigg)\right]\cos( u_1- u_3)\cos u_2\\
&+\left[\frac2{ u_2}-\frac2{ u_2^3}-\frac12{\rm Si}(2 u_2)\right]\cos( u_1- u_3)\sin u_2\\
&+\left[\frac12\gamma_E-\frac43+\frac2{ u_3^2}-\frac12\bigg({\rm Ci}(2 u_3)-\ln(2 u_3)\bigg)\right]\cos( u_1- u_2)\cos u_3\\
&+\left[\frac2{ u_3}-\frac2{ u_3^3}-\frac12{\rm Si}(2 u_3)\right]\cos( u_1- u_2)\sin u_3+\frac1{4\pi}\int_{-1}^1d\mu_2\int_0^{2\pi}d\epsilon\left\{\frac{\mu_1^4}{(1-\mu_1^2)^2}\eta_\zeta^0\right.\\
&\left.+\frac{\mu_1^2\mu_2^2}{(1-\mu_1^2)(1-\mu_2^2)}\eta_\zeta^1+\frac{\mu_1^2\mu_3^2}{(1-\mu_1^2)(1-\mu_3^2)}\eta_\zeta^2+\frac{\mu_2^2\mu_3^2}{(1-\mu_2^2)(1-\mu_3^2)}\eta_\zeta^3\right\}
\end{split}
\end{equation}
where

\begin{gather}
\begin{split}
\eta_\zeta^0=&(1-\mu_1^2)[\cos u_1-\cos( u_1\mu_1)]\cos( u_1\mu_1- u_2+ u_2\mu_2+ u_3- u_3\mu_3)
\end{split}\\
\begin{split}
\eta_\zeta^1=&2(1+\mu_1)\sin\frac12( u_1- u_1\mu_1)\left[(1+\mu_2)\sin\frac12( u_2- u_2\mu_2)\cos\frac12( u_1- u_1\mu_1\right.\\
&\left.+ u_2- u_2\mu_2-2 u_3+2 u_3\mu_3)-(1-\mu_2)\sin\frac12( u_2+ u_2\mu_2)\cos\frac12( u_1+ u_1\mu_1\right.\\
&\left.- u_2+ u_2\mu_2-2 u_3\mu_3)\right]-2(1-\mu_1)\sin\frac12( u_1+ u_1\mu_1)\left[(1+\mu_2)\sin\frac12( u_2\right.\\
&- u_2\mu_2)\cos\frac12( u_1+ u_1\mu_1- u_2+ u_2\mu_2)-(1-\mu_2)\sin\frac12( u_2+ u_2\mu_2)\\
&\left.\times\cos\frac12( u_1- u_1\mu_1+ u_2- u_2\mu_2-2 u_3)\right]
\end{split}\\
\begin{split}
\eta_\zeta^2=&2(1-\mu_1)\sin\frac12( u_1+ u_1\mu_1)\left[(1+\mu_3)\sin\frac12( u_3- u_3\mu_3)\cos\frac12( u_1+ u_1\mu_1+2 u_2\mu_2\right.\\
&\left.-2 u_2+ u_3- u_3\mu_3)-(1-\mu_3)\sin\frac12( u_3+ u_3\mu_3)\cos\frac12( u_1- u_1\mu_1-2 u_2\mu_2- u_3+\right.\\
&\left. u_3\mu_3)\right]-2(1+\mu_1)\sin\frac12( u_1- u_1\mu_1)\left[(1+\mu_3)\sin\frac12( u_3- u_3\mu_3)\cos\frac12( u_1- u_1\mu_1\right.\\
&\left.- u_3+ u_3\mu_3)-(1-\mu_3)\sin\frac12( u_3+ u_3\mu_3)\cos\frac12( u_1+ u_1\mu_1-2 u_2+ u_3- u_3\mu_3)\right]
\end{split}\\
\begin{split}
\eta_\zeta^3=&2(1-\mu_2)\sin\frac12( u_2+ u_2\mu_2)\left[(1+\mu_3)\sin\frac12( u_3- u_3\mu_3)\cos\frac12(2 u_1- u_2+ u_2\mu_2\right.\\
&\left.- u_3- u_3\mu_3)-(1-\mu_3)\sin\frac12( u_3+ u_3\mu_3)\cos\frac12( u_2+ u_2\mu_2- u_3- u_3\mu_3)\right]-\\
&2(1+\mu_2)\sin\frac12( u_2- u_2\mu_2)\left[(1+\mu_3)\sin\frac12( u_3- u_3\mu_3)\cos\frac12( u_2- u_2\mu_2- u_3\right.\\
&\left.+ u_3\mu_3)-(1-\mu_3)\sin\frac12( u_3+ u_3\mu_3)\cos\frac12(2 u_1- u_2- u_2\mu_2- u_3+ u_3\mu_3)\right]
\end{split}
\end{gather}

%

\end{document}